\documentclass[VANCOUVER,STIX1COL]{WileyNJD-v2}
\articletype{Article Type}%

\usepackage[utf8]{inputenc}
\usepackage{dsfont} 

\usepackage{amssymb}
\usepackage{enumitem}

 \definecolor{mDarkBrown}{HTML}{604c38}
 \definecolor{mDarkTeal}{HTML}{23373b}
 \definecolor{mLightBrown}{HTML}{EB811B}
 \definecolor{mLightGreen}{HTML}{14B03D}
 \definecolor{mAlert}{HTML}{AD003D}
 \definecolor{mExample}{HTML}{005580}

\definecolor{UCLwhite}{RGB}{255, 252, 247}
\definecolor{UCLblack}{RGB}{40, 54, 66}
\definecolor{UCLblue}{RGB}{128, 155, 200}
 \colorlet{UCLorange}{mLightBrown}
\definecolor{UCLred}{RGB}{255,102,102}
\definecolor{UCLyellow}{RGB}{255,255,204}
\definecolor{UCLgreen}{RGB}{80,150,0}
\definecolor{UCLdarkBlue}{RGB}{82, 100, 128}
\colorlet{UCLdarkRed}{UCLblack!40!UCLred}
\colorlet{UCLlightOrange}{UCLwhite!85!UCLorange}
\colorlet{UCLlightGreen}{UCLwhite!85!UCLgreen}
\colorlet{UCLlightBlue}{UCLwhite!85!UCLblue}
\colorlet{UCLgrey}{white!65!UCLblack}
\colorlet{UCLlightGrey}{white!85!UCLblack}
\colorlet{UCLdarkGrey}{UCLwhite!30!UCLblack}

\hypersetup{
    pdftitle={One machine, one minute, three billion tetrahedra},    
    pdfauthor={Célestin Marot, Jeanne Pellerin, Jean-François Remacle},     
    pdfsubject={Parallel Delaunay},   
    pdfcreator={Célestin Marot},   
    pdfkeywords={Delaunay, mesh, mesh generation, tetrahedral mesh, parallel, multithread}, 
}

\newcommand{\pointSet}{\ensuremath{S}}
\newcommand{\DT}{\ensuremath{\mbox{DT}}}
\newcommand{\reals}{\ensuremath{\mathds{R}}}

\lstdefinestyle{codestyle}{ 
basicstyle=\fontsize{9pt}{0.8em}\ttfamily\color{UCLblack}, 
breakatwhitespace=false, 
breaklines=true, 
captionpos=b, 
commentstyle=\usefont{T1}{pcr}{bm}{sl}\color{UCLgreen}, 
deletekeywords={}, 
escapeinside={*@}{@*}, 
firstnumber=1, 
frame=single, 
morekeywords=[1]{uint16_t,uint32_t,uint64_t, mesh_t, vertex_t, bnd_t},
keywordstyle=[1]\bfseries\color{UCLdarkBlue}, 
morecomment=[l][\bfseries\color{UCLdarkRed}]{\#},
numbers=left, 
numbersep=10pt, 
numberstyle=\tiny\color{UCLdarkGrey}, 
rulecolor=\color{UCLblack}, 
showstringspaces=false, 
showtabs=false, 
stepnumber=5, 
stringstyle=\color{orange!85!black}, 
tabsize=2, 
}

\usepackage{array}
\newcolumntype{L}[1]{>{\raggedright\let\newline\\\arraybackslash\hspace{0pt}}m{#1}}
\newcolumntype{C}[1]{>{\centering\let\newline\\\arraybackslash\hspace{0pt}}m{#1}}
\newcolumntype{R}[1]{>{\raggedleft\let\newline\\\arraybackslash\hspace{0pt}}m{#1}}

\usepackage{subcaption}
\usepackage{tikz}
\usepackage{pgfplots}

\pgfplotsset{compat=1.13}

\pgfplotscreateplotcyclelist{UCLpgfList}{
UCLdarkBlue,                 thick, every mark/.append style={solid,scale=0.5},  mark=*\\
UCLorange,   densely dashed, thick, every mark/.append style={solid,scale=0.5}, mark=triangle*\\
UCLgreen,    dashdotted,     thick, every mark/.append style={solid,scale=0.5},  mark=square*\\
UCLdarkRed,  densely dotted, thick, every mark/.append style={solid,scale=0.6, fill=UCLdarkRed!50!black}, mark=diamond*\\
}

\pgfplotsset{
    plot coordinates/math parser=false,
    log x ticks with fixed point/.style={ 
        xticklabel={
            \pgfkeys{/pgf/fpu=true}
            \pgfmathparse{exp(\tick)}%
            \pgfmathprintnumber[fixed relative, precision=3]{\pgfmathresult}
            \pgfkeys{/pgf/fpu=false}
        }
    },
    log y ticks with fixed point/.style={
        yticklabel={
            \pgfkeys{/pgf/fpu=true}
            \pgfmathparse{exp(\tick)}%
            \pgfmathprintnumber[fixed relative, precision=3]{\pgfmathresult}
            \pgfkeys{/pgf/fpu=false}
        }
    },
    legend style={font=\footnotesize},
    tick label style={font=\footnotesize},
    label style={font=\fontsize{9}{10.5}\selectfont},
    y label style={at={(-0.065,0.5)}},
    legend style={at={(0.015,0.985)}, anchor=north west, legend cell align=left,align=left, draw=UCLblack},
    xminorticks=false,
    yminorticks=true,
    major grid style={UCLgrey},
    xmajorgrids,
    ymajorgrids,
    minor grid style={UCLlightGrey},
    xminorgrids,
    yminorgrids,
    axis background/.style={fill=white},
    cycle list name=UCLpgfList
}

\newlength\figureheight
\newlength\figurewidth

\begin{document}


\title{One machine, one minute, three billion tetrahedra}

\author[1]{C\'elestin Marot*}
\author[1]{Jeanne Pellerin}
\author[1]{Jean-Fran\c cois Remacle}

\address[1]{Universit\'e catholique de Louvain, iMMC, Avenue Georges Lemaitre 4, bte L4.05.02, 1348 
Louvain-la-Neuve, Belgium}

\abstract[Summary]{

This paper presents a new scalable parallelization scheme 
to generate the 3D Delaunay triangulation of a given set of points.
Our first contribution is an efficient serial implementation of the 
incremental Delaunay insertion algorithm.
A simple dedicated data structure, an efficient sorting of the points and the optimization of the insertion algorithm 
have permitted to accelerate reference implementations by a factor three. 
Our second contribution is a multi-threaded version of
the Delaunay kernel that is able to concurrently insert vertices. 
Moore curve coordinates are used to partition the point set, 
avoiding heavy synchronization overheads.
Conflicts are managed by modifying the partitions
with a simple rescaling of the space-filling curve. 
The performances of our implementation have been measured
on three different processors, 
an Intel core-i7, an Intel Xeon Phi and an AMD EPYC,
on which we have been able to compute 
3 billion tetrahedra in $53$ seconds.
This corresponds to a generation rate of over $55$ million tetrahedra per second.
We finally show how this very efficient parallel Delaunay triangulation can be 
integrated in a Delaunay refinement mesh generator 
which takes as input the triangulated surface boundary of the volume to mesh.
}


\corres{*Corresponding author: \email{celestin.marot@uclouvain.be}}


\maketitle

\section{Introduction} \label{sec:intro}

The Delaunay triangulation is a fundamental geometrical object that associates a unique triangulation to a given point-set in general position. This triangulation and its dual, the Voronoi diagram, have locality properties
that make them ubiquitous in various domains\cite{aurenhammer_voronoi_1991}: mesh generation\cite{cheng_delaunay_2013},
surface reconstruction from 3D scanners point clouds\cite{berger_survey_2017}, astrophysics\cite{cautun_dtfe_2011, ramella_finding_2001}, terrain modeling\cite{heckbert1995fast} etc.
Delaunay triangulation algorithms now face a major challenge: the availability of more and more massive point sets. 
LiDAR or other photogrammetry technologies are used to survey the surface of entire cities and even countries, like the Netherlands\cite{martinez-rubi_taming_2016}.
The size of finite element mesh is also growing with the increased availability of massively parallel numerical solvers.
It is now common to deal with meshes of over several hundred millions of tetrahedra\cite{rasquin_scalable_2014, ibanez_pumi:_2016, vazquez_alya:_2016}.
Parallelizing 3D Delaunay triangulations and Delaunay-based meshing algorithms is however very challenging and the mesh generation process is nowadays considered 
a technological bottleneck in computational engineering \cite{slotnick_cfd_2014}. 

The most part of today's clusters have two levels of parallelism. Distributed memory systems contain thousands of nodes, 
each node being itself a shared memory system with multiple cores.
In recent years, nodes have seen their number of cores and the size of their memory increase,
with some clusters already featuring 256-core processors and up to $12$\textsc{tb} of RAM\cite{Nystrom:2015:BUF:2792745.2792775, fu2016sunway}.
As many-core shared memory machines are becoming standard, 
Delaunay triangulation algorithms designed for shared memory should not only scale well up to 8 cores but to several hundred cores.
In this paper, we complete the approach presented in our previous research note\cite{marot_toward_2017} and 
show that a billion tetrahedra can be computed very efficiently 
on a single many-core shared memory machine.

Our first contribution is a sequential implementation 
of the Delaunay triangulation algorithm in 3D
that is able to triangulate a million points in about $2$ seconds. In comparison, the fastest 3D open-source sequential programs (to our best knowledge): Tetgen\cite{DBLP:journals/toms/Si15}, CGAL\cite{boissonnat2000triangulations} and Geogram\cite{levy2015geogram} all triangulate a million points in about $6$ seconds on one core of a high-end laptop.
Our implementation is also based on the incremental Delaunay 
insertion algorithm,
but the gain in performance is to ascribe 
to the details of the specific data structures we have developed, 
to the optimization of geometric predicates evaluation, and 
to specialized adjacency computations (Section~\ref{sec:serial}).

Our second contribution is a scalable parallel version 
of the Delaunay triangulation algorithm
devoid of heavy synchronization overheads (Section~\ref{sec:parallel}).
The domain is partitioned using the Hilbert curve
and conflicts are detected with a simple coloring scheme.
The performances and scalability are demonstrated on three different machines: 
a high-end four core laptop, 
a 64-core Intel$^\circledR$ Xeon Phi Knight's Landing,
and a recent AMD$^\circledR$ EPYC 64-core machine (Section~\ref{sec:parrallel_performances}).  
On the latter computer, we have been able to generate 
three billion tetrahedra in less than a minute (about $10^7$ points per second).

We finally demonstrate how this efficient Delaunay triangulation algorithm can be 
easily integrated in a tetrahedral mesh generation process where the input 
is the boundary surfaces of the domain to mesh (Section~\ref{sec:mesh_generation}). 

Our reference implementation is open-source and available in Gmsh 4 at \url{http://gmsh.info}.

\section{Sequential Delaunay}\label{sec:serial}

The Delaunay triangulation $\DT(S)$ of a point set $\pointSet$ 
has the fundamental geometrical property 
that the circumsphere of any tetrahedron contains no other point of $\pointSet$ 
than those of the considered tetrahedron. 
%
More formally, a triangulation\footnote{This paper is about 3D meshing. 
Still we use the generic term triangulation instead of tetrahedralization.} 
$T(\pointSet)$ of the $n$ points $\pointSet = \{p_1, \dots , p_n \} \in \reals^3$ 
is a set of non overlapping tetrahedra that covers exactly 
the convex hull $\Omega(\pointSet)$ of the point set,
and leaves no point  $p_i$ isolated.
If the empty circumsphere condition is verified for all tetrahedra, 
the triangulation $T(\pointSet)$ is said to be a Delaunay triangulation.
If, additionally, $\pointSet$ contains no group of 4 coplanar points
and no group of 5 cospherical points,  
then it is said to be in general position, 
and the Delaunay triangulation  $\DT(\pointSet)$ is unique.

The fastest serial algorithm to build the 3D Delaunay triangulation $\DT(\pointSet)$ is probably the Bowyer-Watson algorithm, which
works by incremental insertion of points in the triangulation.
The Bowyer-Watson algorithm, presented in (\S\ref{sec:bowyer_algo}), was devised independently by Bowyer and Watson in 1981 \cite{bowyer_computing_1981, watson_computing_1981}.
Efficient open-source implementations are available: 
Tetgen \cite{DBLP:journals/toms/Si15}, 
CGAL \cite{boissonnat2000triangulations} and 
Geogram \cite{levy2015geogram}.
They are designed similarly and offer therefore similar performances
(Table~\ref{table:profiling_sequential}).

\begin{table}[b]
    \centering
    \begin{tabular}{l r r r r}
    \toprule
       & Ours & Geogram & TetGen & CGAL\\ 
     \midrule
    \textsc{\bfseries Sequential\_Delaunay} & $\mathbf{12.7}$ & $\mathbf{34.6}$ & $\mathbf{32.9}$ & $\mathbf{33.8}$ \\[0.1cm]
    \hspace{0.5cm} \textsc{Init + Sort}& $0.5$ & $4.2$ & $2.1$ & $1.3$\\[0.1cm]
    \hspace{0.5cm} \textsc{Incremental insertion} & $12.2$ & $30.4$ & $30.8$ & $32.5$\\[0.05cm]
    \hspace{1cm} \textsc{Walk}         & $1.0$ & $2.1$  & $ 1.6$ & $1.4$\\[-0.07cm]
    \hspace{1.5cm} \texttt{orient3d}   & $0.7$ & $1.4$ & $1.1$ & $\approx 0.5$\\[0.05cm]
    \hspace{1cm} \textsc{Cavity}       & $6.2$ & $11.4$ & $\approx 10$ & $14.9$\\[-0.07cm]
    \hspace{1.5cm} \texttt{inSphere}   &  $3.2$ & $6.2$ & $5.6$ & $10.5$\\[0.05cm]
    \hspace{1cm} \textsc{DelaunayBall} & $4.5$ & $12.4$ & $\approx 15$ & $15.3$\\[-0.07cm]
    \hspace{1.5cm} Computing sub-determinants & $1.3$ & /~~~ & /~~~ & /~~~\\[0.05cm]
    \hspace{1cm} Other operations      & $0.5$ & $4.5$  & $\approx 4$ & $\approx 1$\\[0.07cm]
    \bottomrule \\
    \end{tabular}
    \caption{Timings for the different steps of the Delaunay incremental insertion (Algorithm~\ref{algo:seqDelaunay}) 
    for four implementations: Ours, Geogram\cite{levy2015geogram}, Tetgen\cite{DBLP:journals/toms/Si15} and CGAL\cite{boissonnat2000triangulations}.    
    Timings in seconds are given for 5 million points (random uniform distribution). 
    The $\approx$ prefix indicates that no accurate timing is available. 
     }
    \label{table:profiling_sequential}
\end{table}

In the remaining of this section, the incremental insertion algorithm is recalled 
and we describe the dedicated data structures that we developed 
as well as the algorithmic optimizations that make our sequential implementation 
three times faster than reference ones.

\subsection{Algorithm overview}\label{sec:bowyer_algo}  

\begin{figure}
    \centering
    \begin{subfigure}[t]{0.4\textwidth}
        \centering
        \includegraphics[width=0.9\textwidth]{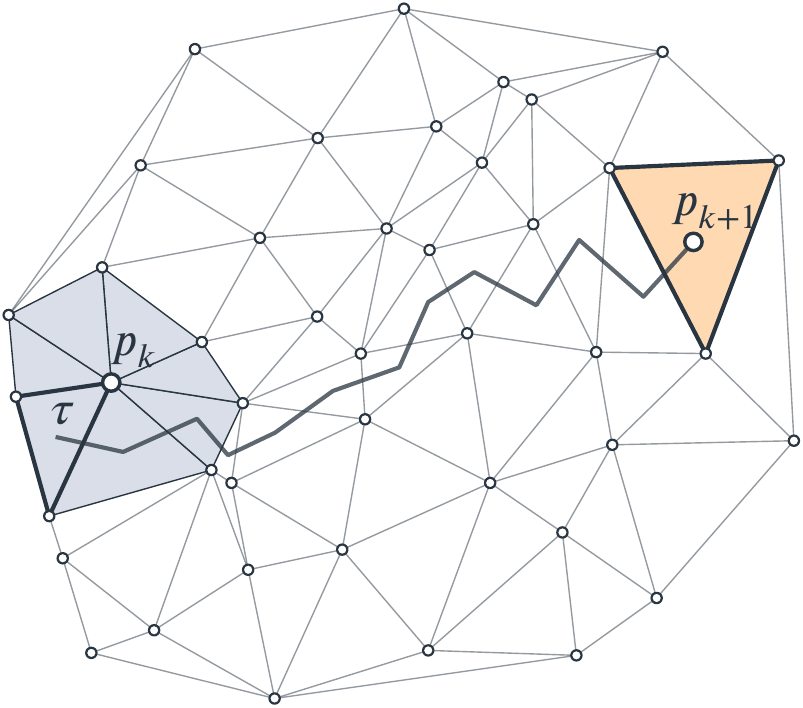}
        \caption{\textsc{Walk}}
        \label{fig:walk}
    \end{subfigure}
    \hspace{0.02\textwidth}
    \begin{subfigure}[t]{0.27\textwidth}
        \centering
        \includegraphics[width=0.9\textwidth]{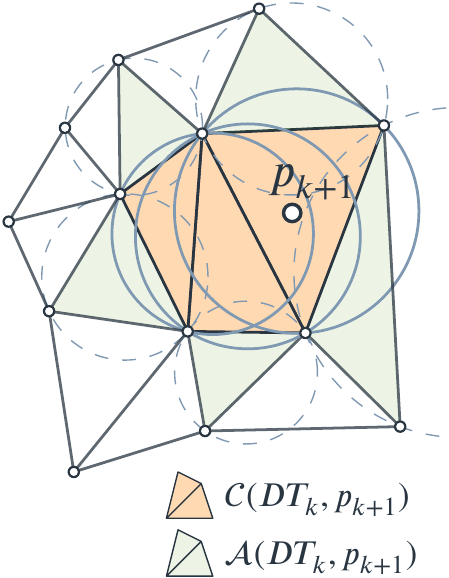}
        \caption{\textsc{Cavity}}
        \label{fig:cavity}
    \end{subfigure}
    \begin{subfigure}[t]{0.27\textwidth}
        \centering
        \includegraphics[width=0.9\textwidth]{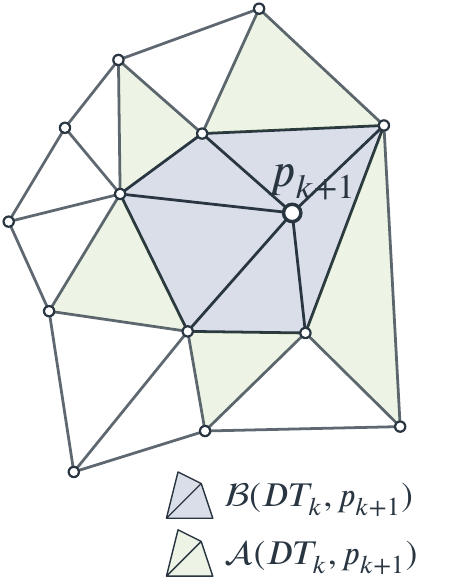}
        \caption{\textsc{DelaunayBall}}
        \label{fig:delaunayball}
    \end{subfigure}
    \caption{Insertion of a vertex $p_{k+1}$ in the Delaunay triangulation $DT_k$.
    \textbf{(a)}  The triangle containing $p_{k+1}$ is obtained by walking toward $p_{k+1}$.
    The {\sc Walk}  starts from $\tau \in {\mathcal{B}}_{p_k}$.
    \textbf{(b)}The {\sc Cavity} function  finds all cavity triangles (orange) whose circumcircle contains the vertex $p_{k+1}$. 
    They are deleted, while cavity adjacent triangles (green) are kept.
    \textbf{(c)} The {\sc DelaunayBall} function creates new triangles (blue) by connecting $p_{k+1}$ to the edges of the cavity boundary.}
    \label{fig:cavity_and_boundary}
\end{figure}

Let ${\DT}_k$ be the Delaunay triangulation of the  subset
$\pointSet_k = \{p_1,\dots,p_k\} \subset S$.  
The \emph{Delaunay kernel} is the procedure
to insert a new point ${p}_{k+1} \in \Omega(\pointSet_k)$ into ${\DT}_k$, 
and construct a new valid Delaunay triangulation $\DT_{k+1}$ 
of $\pointSet_{k+1} = \{p_1,\dots,p_k,p_{k+1}\}.$ 
The \emph{Delaunay kernel}
can be written in the following abstract manner:
\begin{equation}
  {\DT}_{k+1} \leftarrow {\DT}_{k} - {\mathcal C}({\DT}_k,{p}_{k+1}) +
  {\mathcal B}(\DT_k,p_{k+1}) ,\label{eq:delker}
\end{equation}
where the Delaunay cavity ${\mathcal C}({\DT}_k,{p}_{k+1}) $ is the set of
all tetrahedra whose circumsphere contains $p_{k+1}$  (Figure~\ref{fig:cavity}),
whereas the Delaunay ball ${\mathcal B}(\DT_k,p_{k+1})$ is
 a set of tetrahedra filling up the polyhedral hole 
obtained by removing the Delaunay cavity ${\mathcal C}({\DT}_k,{p}_{k+1})$ 
from $\DT_{k}$ (Figure~\ref{fig:delaunayball}).

The complete state-of-the-art incremental insertion algorithm (Algorithm~\ref{algo:seqDelaunay}) 
that we implemented has five main steps that are described in the following.

\begin{algorithm}[b]
\caption{Sequential computation of the Delaunay triangulation DT of a set of vertices $\pointSet$} 
\label{algo:seqDelaunay}
\textbf{Input:} $\pointSet$\\
\textbf{Output:} $\DT(\pointSet)$                                                           \Comment Section~\ref{sec:datastructure}
\begin{algorithmic}[1]
    \Function{Sequential\_Delaunay}{$\pointSet$}
        \State $\tau \gets$ \Call{Init}{$\pointSet$}                       {\color{black!70}\Comment $\tau$ is the current tetrahedron }
        \State $\DT \gets \tau$
        \State $\pointSet' \gets$ \Call{Sort}{$\pointSet \setminus \tau$}                   \Comment Section~\ref{sec:spatial_sorting} 

        \ForAll {$p \in \pointSet'$}
            \State $\tau \gets $ \Call{Walk}{$\DT,\tau,p$}                    

            \State ${\mathcal  C} \gets$ \Call{Cavity}{$\DT,\tau,p$}                        \Comment Section~\ref{sec:cavity}
            \State $\DT \gets \DT \setminus {\mathcal  C}$

            \State ${\mathcal  B} \gets$ \Call{DelaunayBall}{${\mathcal  C},p$}             \Comment Section~\ref{sec:ball}
            \State $\DT \gets \DT \cup {\mathcal  B}$
            \State $\tau \gets$  t $\in {\mathcal  B}$
        \EndFor

        \State \Return $\DT$
    \EndFunction
\end{algorithmic}
\end{algorithm}

\paragraph{ {\sc Init} }

The triangulation is initialized with the tetrahedron formed 
by the first four non-coplanar vertices of the point set $\pointSet$. 
These vertices define a tetrahedron $\tau$ with a positive volume.

\paragraph{ {\sc Sort} }

Before starting point insertion, the points are sorted so that 
two points that have close indices are close in space.
Used alone, this order would result in cavities ${\mathcal C}({\DT}_k,{p}_{k+1})$ containing a pathologically large number of tetrahedra.
Insertions are organized in different randomized stages to avoid this issue\cite{DBLP:conf/compgeom/AmentaCR03}.
The three kernel functions {\sc Walk}, {\sc Cavity} and {\sc DelaunayBall} 
thereby have a constant complexity in practice.
We have implemented a very fast sorting procedure (Section~\ref{sec:spatial_sorting}).

\paragraph{ {\sc Walk}} 

The goal of this step is to identify the tetrahedron $\tau_{k+1}$ 
enclosing the next point to insert $p_{k+1}$.
The search starts from a tetrahedron $\tau_{k}$ in the last Delaunay ball ${\mathcal B}(\DT_k,p_k)$,
and walks through the current triangulation $\DT_k$
in direction of  $p_{k+1}$ (Figure~\ref{fig:walk}).
We say that a point is visible from a facet 
when the tetrahedron defined by this facet and this point has negative volume. 
The \textsc{Walk} function thus iterates on the four facets of $\tau$,
selects one from which the point $p_{k+1}$ is visible, 
and then walks across this facet to the adjacent tetrahedron.
This new tetrahedron is called $\tau$ 
and the \textsc{Walk} process is repeated 
until none of the facets of $\tau$ sees $p_{k+1}$, 
which is equivalent to say that $p_{k+1}$ 
is inside $\tau$ (see Figure \ref{fig:walk}). 

The visibility walk algorithm is guaranteed to terminate for Delaunay triangulations\citep{DeFloriani1991}.
If the points have been sorted, 
the number of walking steps is essentially constant\cite{DBLP:journals/cad/Remacle17}. 
Our implementation of this robust and efficient walking algorithm 
is similar to other available implementations. 

\paragraph{ {\sc Cavity} }

Once the tetrahedron $\tau \gets \tau_{k+1}$ that contains the point to insert  $p_{k+1}$ 
has been identified, the function \textsc{Cavity} finds all
tetrahedra whose circumsphere contain $p_{k+1}$ and deletes them.
The Delaunay cavity ${\mathcal C}({\DT}_k,{p}_{k+1})$ is simply connected
and contains $\tau$ \cite{shewchuk1997delaunay}, it is then built using a breadth-first search algorithm.
The core and most expensive operation of  the \textsc{Cavity} function 
is the \texttt{inSphere} predicate,
which evaluates whether a point $e$ is inside/on or outside the circumsphere of given tetrahedron. 
This \textsc{Cavity}  function is thus an expensive function of the incremental insertion, 
which accounts for about 33\% of the total computation time (Table~\ref{table:profiling_sequential}).
To accelerate this, we propose in Section~\ref{sec:cavity} 
to precompute sub-components of the \texttt{inSphere} predicate.

\paragraph{ {\sc DelaunayBall} }

Once the cavity has been carved,
the \textsc{DelaunayBall} function
first generates  a set of new tetrahedra 
adjacent to the newly inserted point $p_{k+1}$ and filling up the cavity,
and then updates the mesh structure.
In particular, the mesh update consists 
in the computation of adjacencies between the newly created tetrahedra.
This is the most expensive step of the algorithm, 
with about 40\% of the total computation time (Table~\ref{table:profiling_sequential}).
To accelerate this step, we replace general purpose elementary operations
like tetrahedron creation/deletion or adjacency computation,
with batches of optimized operations making benefit from
a cavity-specific data structure (Section~\ref{sec:ball}).

\subsection{Mesh data structure}\label{sec:datastructure}

One key for enhancing performances of a Delaunay triangulation algorithm 
resides in the optimization of the data structure used to store the triangulation. 
Various data structure designs have been proposed that are very flexible
and allow representing hybrid meshes, high order meshes, add mesh elements of any type, 
or manage adjacencies around vertices, edges, etc.
The versatility of such general purpose data structures has a cost,
both in terms of storage and efficiency. 
Here, our aim is to have a structure as lightweight and fast as possible, 
dealing exclusively with 3D triangulations. 
Our implementation is coded in plain {\tt C} language, 
with arrays of doubles, floats, and integers to store mesh topology and geometry.
This seemingly old-style coding has important advantages in terms of optimization and 
parallelization because it compels us to use simple and straightforward algorithms.

\begin{lstlisting}[caption={The aligned mesh data structure {\tt mesh\_t} we use to store the vertices and tetrahedra of a Delaunay triangulation in 3D.}, label=mesh_struct, captionpos=b, float]
typedef struct {
    double coordinates[3];
    uint64_t padding;
} point3d_t;

typedef struct {
    struct {
        uint32_t* vertex_ID;
        uint64_t* neighbor_ID;
        double* sub_determinant;     
        uint64_t num;            // number of tetrahedra
        uint64_t allocated_num;  // capacity [in tetrahedra]
    } tetrahedra;

    struct {
        point3d_t* vertex;
        uint32_t num;           // number of vertices
        uint32_t allocated_num; // capacity [in vertices]
    } vertices;
} mesh_t;
\end{lstlisting}

\begin{figure}
\centering
\begin{subfigure}[h]{0.35\textwidth}
    \centering
    \includegraphics[width=\textwidth]{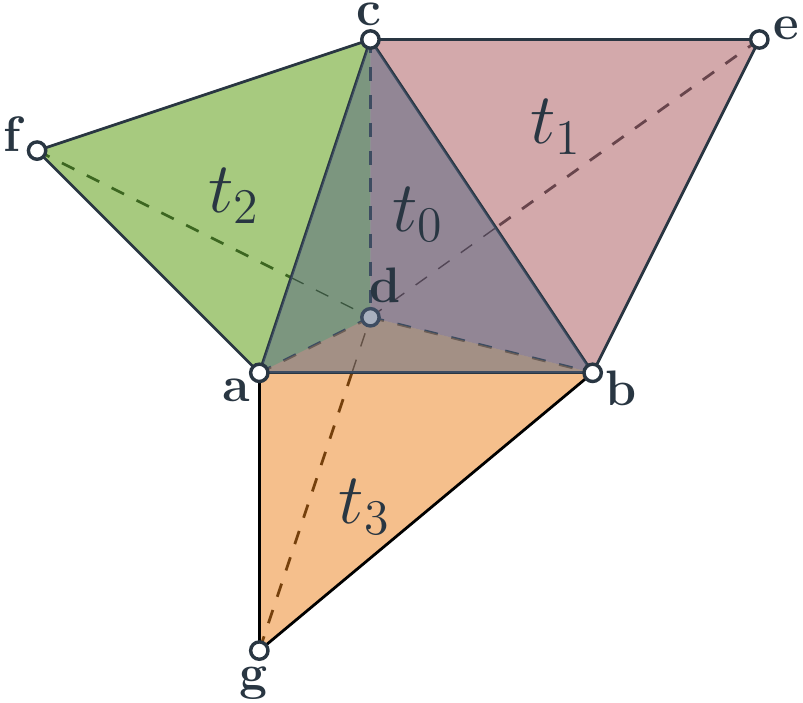}
    \vspace{1cm}
\end{subfigure}
\hspace{0.05\textwidth}
\begin{subfigure}[h]{0.35\textwidth}
	\centering
	\footnotesize
	\begin{equation*}
		\begin{matrix}
		\toprule
		\text{memory index} & ~~~\texttt{vertex\_ID}~~~ & \texttt{neighbor\_ID}\\\midrule
		4 t_0 & a & 4 t_1+3\\
		4 t_0 + 1 & b & 4 t_2+3\\
		4 t_0 + 2 & c & 4 t_3+3\\
		4 t_0 + 3 & d & -\\
		: & & \\
		4 t_1 & b & -\\
		4 t_1 + 1 & c & -\\
		4 t_1 + 2 &d & -\\
		4 t_1 + 3 &e & 4 t_0 + 0\\
		: & & \\
		4 t_2 &a & -\\
		4 t_2 + 1 &d & -\\
		4 t_2 + 2 &c & -\\
		4 t_2 + 3 &f & 4 t_0 + 1\\
		: & & \\
		4 t_3 & a & -\\
		4 t_3 + 1 & b & -\\
		4 t_3 + 2 & d & -\\
		4 t_3 + 3 & g & 4 t_0 + 2\\
		\bottomrule
		\end{matrix}
	\end{equation*}
	\vspace{0.2cm}
\end{subfigure}
    \caption{Four adjacent tetrahedra : $t_0,t_1,t_2,t_3$ and one of their possible memory representations in the \texttt{tetrahedra} data structure given in Listing \ref{mesh_struct}. \texttt{tetrahedra.neighbor\_ID$[4 t_i + j]/4$} gives the index of the adjacent tetrahedron opposite to \texttt{tetrahedra.vertex\_ID$[4 t_i + j]$} in the tetrahedron \texttt{$t_i$} and \texttt{tetrahedra.neighbor\_ID$[4 t_i + j]$} gives the index where the inverse adjacency is stored.}
    \label{fig:data_struct}
\end{figure}

The mesh data only contains vertices and tetrahedra explicitly,
and all topological and geometrical information can be deduced from it.
However, in order to speed up  mesh operations, 
it is beneficial to store additional connectivity information. 
A careful trade-off needs however to be made
between computational time and memory space. 
The only connectivity information we have chosen to store 
is the adjacency between tetrahedra,
as this allows walking through the mesh using local queries only.

\subparagraph{Vertices}
Vertices are stored in a single array of structures {\tt point3d\_t} 
(see Listing \ref{mesh_struct}).
For each vertex, in addition to the vertex coordinates, 
a {\tt padding} variable is used to align the structure to 32 bytes
(3 doubles of 8 bytes each, and an additional padding variable of 8 bytes
sum up to a structure of 32 bytes) 
and conveniently store temporarily some auxiliary vertex related values 
at different stages of the algorithm. 
Memory alignment ensures that a vertex does not overlap 
two cache lines during memory transfer.
Modern computers usually work with cache lines of 64 bytes.
The padding variable in the vertex structure ensures that 
a vertex is always loaded in one single memory fetch. 
Moreover, 
aligned memory allows to take advantage 
of the vectorization capabilities of modern microprocessors. 

\subparagraph{Tetrahedra}
Each tetrahedron knows about its $4$
vertices and its $4$ neighboring tetrahedra.
These two types of adjacencies are stored in separate arrays.
The main motivation for this storage is flexibility and, once more, memory alignment.
Keeping good memory alignment properties on a tetrahedron structure evolving with the implementation is cumbersome.
In addition, it provides little to no performance gain in this case.
On the other hand with parallel arrays, additional information per tetrahedron  (e.g. a color for each tetrahedron, sub-determinants etc.)
can be added easily without disrupting memory layout.
%
Each tetrahedron is identified by the indices of its four vertices in 
the {\tt vertices} structure. 
Vertex indices of tetrahedron {\tt t} are read between positions {\tt 4*t} and {\tt 4*t+3} 
in the global array {\tt tetrahedra.vertex\_ID} storing all tetrahedron vertices.
Vertices are ordered so that the volume of the tetrahedron is positive. 
An array of double, {\tt sub\_determinant}, is used to store 4 values per tetrahedron.
This space is used to speed up geometric predicate evaluation (see Section~\ref{sec:cavity}).

\subparagraph{Adjacencies}
By convention, the $i$-th facet of a tetrahedron is the facet opposite the $i$-th vertex,
and the $i$-th neighbor is the tetrahedron adjacent to that facet.
In order to travel efficiently through the triangulation,  
facet indices are stored together with the indices 
of the corresponding adjacent tetrahedron,
thanks to an integer division and its modulo. 
The scheme is simple.
Each adjacency is represented by the integer obtained 
by multiplying by four the index of the tetrahedron 
and adding the internal index of the facet in the tetrahedron. 
Take, for instance, two tetrahedra $t_1$ and $t_2$ sharing facet $f$,
whose index is respectively $i_1$ in $t_1$ and $i_2$ in $t_2$.
The adjacency is then represented as 
{\tt tetrahedra.neighbor\_ID[4t$_1$+i$_1$]=4t$_2$+i$_2$}  and  
{\tt tetrahedra.neighbor\_ID[4t$_2$+i$_2$]=4t$_1$+i$_1$}. 
This multiplexing avoids a costly looping over the facets of a tetrahedron.
The fact that it reduces by a factor 4 the maximum number of elements in a mesh
is not a real concern since, element indices and adjacencies being stored as 64 bytes unsigned integers,
the maximal number of element in a mesh is $2^{62} \simeq  4.6~10^{18}$,
which is huge. 
Note also that division and modulo by $4$ are very cheap bitwise operations
for unsigned integers:  $i/4 = i \gg 2$ and $i\%4 = i \& 3$. 

\subparagraph{Memory footprint}
The key to an efficient data structure is the balance 
between its memory footprint and the computational cost of its modification.
We do not expect to generate meshes of more than {\tt UINT32\_MAX} vertices, 
i.e. about 4 billion vertices on one single machine.
Each vertex therefore occupies 32 bytes, 24 bytes for its coordinates and 8 bytes for the padding variable.
On the other hand,  the number of tetrahedra could itself be larger than 4 billion,
so that a 64 bits integer is needed for element indexing.
Each tetrahedron occupies 80 bytes, $4\times 4 = 16$ bytes for the vertices, $4\times 8 = 32$ bytes for the neighbors, 32 bytes again for the sub-determinants.
In average a tetrahedral mesh of $n$ vertices has a little more than $6n$ tetrahedra.
Thus, our mesh data structure requires $\approx 6 \times 80 + 32 = 512$ bytes per vertex.

\subsection{Fast spatial sorting}\label{sec:spatial_sorting}

The complexity of the Delaunay triangulation algorithm depends 
on the number of tetrahedra traversed during the walking steps,
and on the number of tetrahedra in the cavities. 
An appropriate spatial sorting is required to improve locality, 
i.e.,  reduce the walking steps,
while retaining enough randomness 
to have cavity sizes independent of the number of vertices already inserted in the mesh
(by cavity size, we mean the number of tetrahedra in the cavity, not the volume of the cavity).

An efficient spatial sorting scheme has been proposed 
by Boissonnat et al.\citep{Boissonnat:2009:ICD:1542362.1542403} 
and is used in state-of-the-art Delaunay triangulation implementations.
The main idea is to first shuffle the vertices,
and to group them in rounds of increasing size 
(a first round with, e.g., the first 1000 vertices, a second with the next 7000 vertices, etc.)
as described by 
the Biased Randomized Insertion Order (BRIO)\cite{DBLP:conf/compgeom/AmentaCR03}.
Then, each group is ordered along a space-filling curve.
The space-filling curve should have the property 
that successive points on the curve are geometrically close to each other. 
With this spatial sorting, the number of walking steps 
between two successive cavities
remains small and essentially constant\cite{DBLP:journals/cad/Remacle17}.
Additionally, proceeding by successive rounds according to the BRIO algorithm
tends to reduce the average size of cavities.


The Hilbert curve is a continuous self-similar (fractal) curve.
It has the interesting property to be space-filling, i.e., 
to visit exactly once all the cells of a regular grid
with $2^m\times2^m\times2^m$ cells, $m \in \mathbb{N}$. 
A Moore curve is a closed Hilbert curve.
Hilbert and Moore curves have the sought spatial locality properties,
in the sense that points close to each other on the curve are also close to each other in $\mathbb{R}^3$\cite{DBLP:journals/comgeo/HaverkortW10,DBLP:journals/gis/AbelM90}.
Any 3D point set can be ordered by a Hilbert/Moore curve 
according to the order in which the curve visits the grid cells that contains the points. 
The Hilbert/Moore index is thus an integer value $d \in \{0,1,2,\ldots,2^{3m}-1\}$,
and several points might have the same index. 
The bigger $m$ is for a given point set, the smaller the grid cells are,
and hence the lower the probability of having two points with the same Hilbert/Moore index. 
Given a 3D point set with $n$ points, 
there are in average $n/2^{3m}$ points per grid cell. Therefore,
choosing $m=k~log_2(n)$ with $k$ constant
ensures the average number of points in grid cells to be independent of $n$.
If known in advance, the minimum mesh size can also be taken into account: $m$ can be chosen such that a grid cell never contain more than a certain number of vertices, thus capturing non-uniform meshes more precisely.
For Delaunay triangulation purposes, 
the objective is both
to limit the number of points with the same indices (in the same grid cell) and 
to have indices within the smallest range as possible to accelerate subsequent sorting operations. 
There is thus a balance to find, so that Hilbert indices are neither too dense nor too sparse.

Computing the Hilbert/Moore index of one cell is a somewhat technical point. 
The Hilbert/Moore curve is indeed fractal,  
which means recursively composed of replicas of itself 
that have been scaled down by a factor two, 
rotated and optionally reflected. 
Those reflections and rotations can be efficiently computed with bitwise operations.
Various transformations can be applied to point coordinates 
to emulate non-regular grids,
a useful functionality we will resort to in Section~\ref{sec:partitioning}.
Hamilton gives extensive details on how to compute Hilbert indices\cite{DBLP:journals/ipl/HamiltonR08} 
and we provide open-source implementation. 

\begin{figure}[b]
    \centering
    \begin{tikzpicture}
    	\begin{axis}[
        cycle list/.define={my mark}{
	        semithick,every mark/.append style={solid, scale=0.95},mark=*\\
	        semithick,every mark/.append style={solid, scale=1},mark=triangle*\\
	        semithick,every mark/.append style={solid, scale=0.78},mark=square*\\
	        semithick,every mark/.append style={solid, scale=1.15},mark=diamond*\\
	    },
        cycle multi list={
    		my mark
                \nextlist
            UCLpgfList
                \nextlist
            densely dashdotted
                \nextlist
        },
		width=0.8\textwidth,
		height=0.5\textwidth,
		xmode=log,
		xmin=1000,
		xmax=1000000000,
		xlabel={Number of \{key, value\} pairs sorted (uniform distribution of 64-bit keys, 64-bit values)},
		ymode=log,
		ymin=0.0001,
		ymax=2000,
		ylabel={Time [s]},
		log y ticks with fixed point
	]
    	\addplot+[very thick, solid]
table[row sep=crcr]{
10 1.9669535e-06 \\
20 2.5033955e-06 \\
50 4.708767e-06 \\
100 9.953975700000001e-05 \\
200 7.724761975e-05 \\
500 0.000133931637 \\
1000 0.00021713972075 \\
2000 0.00040972232825 \\
5000 0.000872015953 \\
10000 0.0019903182982500003 \\
20000 0.005133509636000001 \\
50000 0.006581246852999999 \\
100000 0.00596576929075 \\
200000 0.0060326457025 \\
500000 0.0076662898065 \\
1000000 0.007896244525750001 \\
2000000 0.010864436626500001 \\
5000000 0.016227245330749998 \\
10000000 0.027702033519749997 \\
20000000 0.05039292573925 \\
50000000 0.14707773923875 \\
100000000 0.30245923995949997 \\
200000000 0.6352652311325 \\
500000000 1.7549694180487498 \\
1000000000 3.5359137058259997 \\
};
\addlegendentry{Ours};

\pgfplotsset{cycle list shift=4}

\addplot
table[row sep=crcr]{
10 2.624932675e-05 \\
20 3.90876085e-05 \\
50 4.1713006750000005e-05 \\
100 4.84734775e-05 \\
200 6.270967425e-05 \\
500 0.00010683108125 \\
1000 0.009804312795249999 \\
2000 0.00647553335875 \\
5000 0.00627842079875 \\
10000 0.00744883902375 \\
20000 0.0067417258395 \\
50000 0.007091197185 \\
100000 0.00763962697225 \\
200000 0.00789789669225 \\
500000 0.01739495620125 \\
1000000 0.01763398200275 \\
2000000 0.031136920675499998 \\
5000000 0.06536783557375 \\
10000000 0.12517647631475 \\
20000000 0.20611622277650002 \\
50000000 0.500213551335 \\
100000000 1.13293639291075 \\
200000000 1.94332559593025 \\
500000000 4.952028019353749 \\
1000000000 10.46564708687375 \\
};
\addlegendentry{GCC libstdc++ parallel mode: \texttt{std::sort()}};

\addplot
table[row sep=crcr]{
10 0.0049452781675 \\
20 0.00463116169 \\
50 0.00439172983175 \\
100 0.004990041256 \\
200 0.004992723465 \\
500 0.005858063698 \\
1000 0.00746041536325 \\
2000 0.0097097158435 \\
5000 0.015000224113750002 \\
10000 0.03722667694075 \\
20000 0.06519442796725 \\
50000 0.29366952180825 \\
100000 0.3608297109605 \\
200000 0.36689651012425 \\
500000 0.3881583213805 \\
1000000 0.37318855524075 \\
2000000 0.37947481870649996 \\
5000000 0.3934483528135 \\
10000000 0.49117553234075007 \\
20000000 0.5680510401725 \\
50000000 0.7443948388102499 \\
100000000 0.9886944293975001 \\
200000000 1.692325472832 \\
500000000 4.3409979939462495 \\
1000000000 9.974849998951 \\
};
\addlegendentry{Intel PSTL: \texttt{std::sort(par\_unseq,..)}};

\addplot
table[row sep=crcr]{
10 2.20537225e-06 \\
20 1.0550021999999999e-05 \\
50 1.5556812e-05 \\
100 2.396106725e-05 \\
200 4.27365305e-05 \\
500 0.006611287593749999 \\
1000 0.006672978401 \\
2000 0.00692230463025 \\
5000 0.00800502300275 \\
10000 0.009573221206500001 \\
20000 0.014009177684750001 \\
50000 0.01634925603875 \\
100000 0.02396374940875 \\
200000 0.034687399864000004 \\
500000 0.072580933571 \\
1000000 0.12891948223125002 \\
2000000 0.21039855480175002 \\
5000000 0.3493477106095 \\
10000000 0.6351625323295 \\
20000000 1.80491423606875 \\
50000000 3.2568714618684997 \\
100000000 6.97924822568875 \\
200000000 16.6836329698565 \\
500000000 34.066753745079 \\
1000000000 71.47431498765926 \\
};
\addlegendentry{Intel TBB: \texttt{parallel\_sort()}};

\addplot
table[row sep=crcr]{
10 0.00026599193633333333 \\
20 0.00020210196566666666 \\
50 0.000206572314 \\
100 0.000226661563 \\
200 0.00022760406133333333 \\
500 0.0002655846376666666 \\
1000 0.00037219623700000007 \\
2000 0.0005821573236666667 \\
5000 0.0010938048363333334 \\
10000 0.0021220693983333336 \\
20000 0.005683961075966666 \\
50000 0.015206115692699997 \\
100000 0.029705364691899997 \\
200000 0.039216440667483324 \\
500000 0.0709302238503 \\
1000000 0.06436710593589999 \\
2000000 0.07847272759948332 \\
5000000 0.14270672133821666 \\
10000000 0.2439478080099833 \\
20000000 1.5062262997650164 \\
50000000 3.2408288223671162 \\
100000000 9.221005310056132 \\
200000000 18.24343003189788 \\
500000000 15.55906099130688 \\
1000000000 25.74380273694795 \\
};
\addlegendentry{Boost: \texttt{block\_indirect\_sort()}};

\addplot
table[row sep=crcr]{
10 0.00027014936033333334 \\
20 0.0002196729183333333 \\
50 0.00022155294799999998 \\
100 0.00024386992066666668 \\
200 0.00027805815133333337 \\
500 0.000337212036 \\
1000 0.000422572096 \\
2000 0.0006343188386666666 \\
5000 0.0013468737403333332 \\
10000 0.00260536621 \\
20000 0.006177924511650001 \\
50000 0.01651811351396667 \\
100000 0.06023133136326666 \\
200000 0.15017890756345 \\
500000 0.16510269356261667 \\
1000000 0.5059628471109999 \\
2000000 0.5399737174933167 \\
5000000 0.632891858866 \\
10000000 0.8206652786582999 \\
20000000 1.0489013134191332 \\
50000000 1.5477667302512996 \\
100000000 2.1086864260956997 \\
200000000 3.057707337849016 \\
500000000 5.780018491235933 \\
1000000000 10.183240465571565 \\
};
\addlegendentry{Boost: \texttt{sample\_sort()}};

\addplot
table[row sep=crcr]{
10 0.00018136079133333335 \\
20 0.0002002952 \\
50 0.00019924218433333332 \\
100 0.00024604797366666666 \\
200 0.0002825371923333333 \\
500 0.00032664214599999994 \\
1000 0.00041477382200000003 \\
2000 0.0006457815563333334 \\
5000 0.0013380882640000001 \\
10000 0.002500207473666667 \\
20000 0.006121237513733332 \\
50000 0.016715214215066665 \\
100000 0.035755309338316665 \\
200000 0.11245743222538333 \\
500000 0.30957256530711663 \\
1000000 0.32618095589185 \\
2000000 1.0459908782195833 \\
5000000 1.21075784098345 \\
10000000 1.4772312861558332 \\
20000000 2.1749242257945665 \\
50000000 3.4510353659587496 \\
100000000 5.31336242898025 \\
200000000 8.777731724580049 \\
500000000 18.96426319908355 \\
1000000000 35.71476886731888 \\
};
\addlegendentry{Boost: \texttt{parallel\_stable\_sort()}};

\pgfplotsset{cycle list shift=8}

\addplot
table[row sep=crcr]{
10 1.3357959999999999e-05 \\
20 2.2562220499999998e-05 \\
50 2.514943475e-05 \\
100 3.6166980750000004e-05 \\
200 4.5879743999999994e-05 \\
500 9.629596025e-05 \\
1000 0.00018430780624999999 \\
2000 0.00036296434724999993 \\
5000 0.0009656855837499999 \\
10000 0.002139898017 \\
20000 0.0048787212 \\
50000 0.0121674984695 \\
100000 0.0257301004605 \\
200000 0.05290616489925 \\
500000 0.14627305511400002 \\
1000000 0.2975456640125 \\
2000000 0.6243239194154999 \\
5000000 1.6524691786617498 \\
10000000 3.46005234029125 \\
20000000 7.238006452098251 \\
50000000 19.23583775665625 \\
100000000 39.85949175897975 \\
200000000 82.7682679425925 \\
500000000 218.26002119667825 \\
1000000000 452.11503954045475 \\
};
\addlegendentry{GCC libstdc++: \texttt{std::sort()}};

\pgfplotsset{cycle list shift=6}

\addplot
table[row sep=crcr]{
10 2.267584225e-05 \\
20 2.493429925e-05 \\
50 3.5770237250000005e-05 \\
100 0.0007813321425 \\
200 0.0005983719602500001 \\
500 0.000790021382 \\
1000 0.00109773408625 \\
2000 0.0017643468455 \\
5000 0.004051278345250001 \\
10000 0.00835262052725 \\
20000 0.016750779934000003 \\
50000 0.044236987829249994 \\
100000 0.0926530146975 \\
200000 0.19222854636599998 \\
500000 0.5090031605212499 \\
1000000 1.0710154697299998 \\
2000000 2.2419430892915 \\
5000000 5.944165723398251 \\
10000000 12.569218661636 \\
20000000 25.97287513315675 \\
50000000 68.39253227319574 \\
100000000 142.82050423510375 \\
200000000 296.257074170746 \\
500000000 778.572088221088 \\
1000000000 1617.8464964209124 \\
};
\addlegendentry{glibc: \texttt{qsort()}};
	\end{axis}
	\end{tikzpicture}
    \caption{Performances of {\tt HXTSort} for sorting $\{key,value\}$ pairs on an Intel$^\circledR$ Xeon Phi$^\text{TM}$ 7210 CPU 
	and comparison with widely used implementations. 
	}
	\label{fig:sort_KNL}
\end{figure}
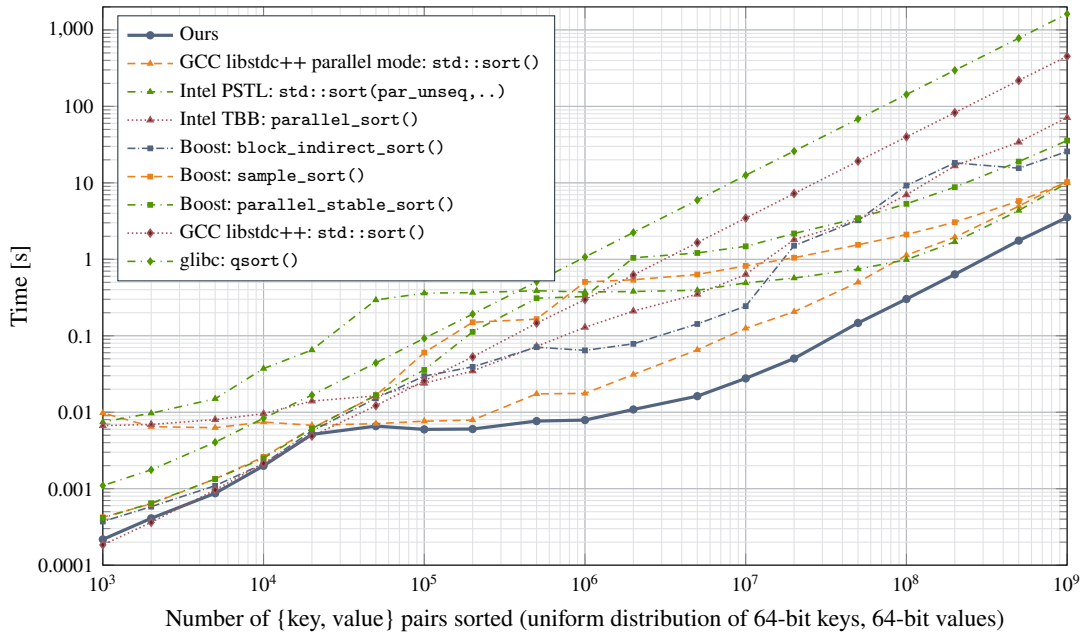

Once the Hilbert/Moore indices have been computed, points can be sorted accordingly
in a data structure where the Hilbert/Moore index is the key, and the point the associated value.
An extremely well suited strategy to sort bounded integer keys is the radix sort%
\cite{DBLP:journals/tc/Blelloch89,DBLP:conf/sc/ZaghaB91,sohn1998load,DBLP:conf/ipps/SatishHG09},
a non-comparative sorting algorithm working digit by digit. 
The base of the digits, the radix, can be freely chosen.
Radix sort has a $O(wn)$ computational complexity, 
where $n$ is the number of $\{key,value\}$ pairs 
and $w$ the number of digits (or bits) of the keys.
In our case, the radix sort has a complexity in $O(m n) = O(n~log(n))$, 
where $n$ is the number of points 
and $m=k~log_2(n)$ the number of levels of the Hilbert/Moore grid. 
In general, $m$ is small because a good resolution of the space-filling curve is not needed. 
Typically, the maximum value among keys is lower than the number of values to be sorted. 
We say that keys are \emph{short}. 
Radix-sort is able to sort such keys extremely quickly.

%
Literature is abundant on parallel radix sorting and impressing performances 
are obtained on many-core CPUs and GPus\cite{10.1007/978-3-642-23397-5_16,DBLP:conf/sigmod/PolychroniouR14,DBLP:conf/sigmod/SatishKCNLKD10,GPU_implementation,bell2011thrust,DBLP:conf/egh/SenguptaHZO07}.
However, implementations are seldom available 
and we were not able to find a parallel radix sort implementation 
properly designed for many-core CPUs.
We implemented \texttt{HXTSort},
that is available independently as open source at \url{https://www.hextreme.eu/hxtsort}. 
Figure \ref{fig:sort_KNL} compares the performances of \texttt{HXTSort} with \texttt{qsort}, 
\texttt{std::sort} and the most efficient implementations that we are aware of for sorting 64-bit key and value pairs.
Our implementation is fully multi-threaded and takes advantage of the
vectorization possibilities offered by modern computers 
such as the AVX512 extensions on the Xeon PHI.
It has been developed primarily for sorting Hilbert indices, which are typically short. 
We use different strategies depending on the key bit size.
These are the reason why \texttt{HXTSort} outperforms the Boost Sort Library, GNU's and Intel's parallel implementation of the standard library and Intel TBB when sorting Hilbert indices.


\subsection{Improving {\sc Cavity}: spending less time in geometric predicates}\label{sec:cavity}

The first main operation of the Delaunay kernel 
is the construction of the cavity ${\mathcal C}({\DT}_k,{p}_{k+1})$ 
formed by all the tetrahedra $\{a,b,c,d\}$ 
whose circumscribed sphere encloses $p_{k+1}$ (Figure~\ref{fig:cavity_and_boundary}).
This step of the incremental insertion represents about one third
of the total execution time in available implementations (Table~\ref{table:profiling_sequential}).
The cavity is initiated with a first tetrahedron $\tau$ containing $p_{k+1}$
determined with the {\sc Walk} function
(Algorithm~\ref{algo:seqDelaunay} and Figure~\ref{fig:walk}),
and then completed by visiting the neighboring tetrahedra 
with a breadth-first search algorithm. 

\paragraph{Optimization of the \texttt{inSphere} predicate}

The most expensive operation of  the \textsc{Cavity} function 
is the fundamental geometrical evaluation of whether a given point $e$ is inside, exactly on, or outside 
the circumsphere of a given tetrahedra $\{a,b,c,d\}$ (Table~\ref{table:profiling_sequential}). 
This is evaluated using the {\tt inSphere} predicate that computes the sign of the following determinant:

\begin{minipage}{0.65\textwidth}
\begin{align*}
    \text{\tt{inSphere}}(a,b,c,d,e) &=
    \begin{vmatrix}
        a_x  \qquad &a_y \qquad & a_z  \qquad & \lVert a \rVert ^2 \qquad & 1 \\
        b_x  \qquad &b_y \qquad & b_z \qquad & \lVert b \rVert ^2 \qquad & 1 \\
        c_x  \qquad &c_y \qquad & c_z  \qquad & \lVert c \rVert ^2  \qquad & 1 \\
        d_x  \qquad &d_y \qquad & d_z \qquad & \lVert d \rVert ^2 \qquad & 1 \\
        e_x  \qquad &e_y \qquad & e_z \qquad & \lVert e \rVert ^2  \qquad & 1 \\
    \end{vmatrix} 
    =
    \begin{vmatrix}
        b_x - a_x \qquad & b_y-a_y  \qquad & b_z - a_z \qquad & \lVert b - a \rVert ^2 \\
        c_x - a_x \qquad & c_y-a_y  \qquad & c_z - a_z \qquad & \lVert c - a \rVert ^2 \\
        d_x - a_x \qquad & d_y-a_y  \qquad & d_z - a_z \qquad & \lVert d - a \rVert ^2 \\
        e_x - a_x \qquad & e_y-a_y  \qquad & e_z - a_z \qquad & \lVert e - a \rVert ^2 \\
    \end{vmatrix} \\
\end{align*}

\end{minipage}
\begin{minipage}{0.25\textwidth}
	\hfill	
    \begin{tikzpicture}[scale=0.5, every node/.style={draw,shape=circle,inner sep=0pt, minimum size=1mm,fill=UCLblack}]
    \node [label={below left:$a$}] (a) at (-1.5,-1.15) {};
    \node [label={below:$b$}] (b) at (-30:2) {};
    \node [label={above:$c$}] (c) at (100:2) {};
    \node [label={above right:$d$}] (d) at (0,-0.352) {};
    \node [label={above:$e$}] (e) at (150:2.2) {};

    \shade[ball color = UCLlightBlue, opacity = 0.4] (0,0) circle (2);
    \draw[color = UCLblue] (0,0) circle (2);
    \draw (b) -- (c) -- (a) -- cycle;
    \draw (b) -- (a);
    \draw[dashed] (d) -- (a);
    \draw[dashed] (d) -- (b);
    \draw[dashed] (c) -- (d);

    \begin{scope}[scale=0.88, shift={(0,-1)}]
    \draw[color = UCLblue] (-2,0) arc (180:360:2 and 0.6);
    \draw[dashed, color = UCLblue] (2,0) arc (0:180:2 and 0.6);
    \end{scope} 
    \end{tikzpicture}

	\vspace{2em}
\end{minipage}

This is a very time consuming computation,
and to make it more efficient, we propose to expand the $4 \times 4$ determinant 
into a linear combination of four $3 \times 3$ determinants independent of point $e$. 

\begin{flalign}
    \hspace{1.5cm}\text{{\tt{inSphere}}}(a,b,c,d,e) = &&-& (e_x - a_x)
                    &\begin{vmatrix}
                      b_y - a_y \qquad & b_z - a_z \qquad & \lVert b - a \rVert ^2 \\
                      c_y - a_y \qquad & c_z - a_z \qquad & \lVert c - a \rVert ^2 \\
                      d_y - a_y \qquad & d_z - a_z \qquad & \lVert d - a \rVert ^2 \\
                    \end{vmatrix} 
                    \nonumber 
	&&+&& (e_y - a_y) &\begin{vmatrix}
                      b_x - a_x \qquad & b_z - a_z \qquad & \lVert b - a \rVert ^2 \\
                      c_x - a_x \qquad & c_z - a_z \qquad & \lVert c - a \rVert ^2 \\
                      d_x - a_x \qquad & d_z - a_z \qquad & \lVert d - a \rVert ^2 \\
                    \end{vmatrix} \hspace{1.5cm}&& \hspace{3cm} \nonumber\\[0.1cm]
    \hspace{1.5cm}&&-& (e_z - a_z)     &\begin{vmatrix}
                      b_x - a_x \qquad & b_y - a_y \qquad & \lVert b - a \rVert ^2 \\
                      c_x - a_x \qquad & c_y - a_y \qquad & \lVert c - a \rVert ^2 \\
                      d_x - a_x \qquad & d_y - a_y \qquad & \lVert d - a \rVert ^2 \\
                    \end{vmatrix} 
					&&+&& \lVert e - a \rVert^2
                    &\begin{vmatrix}
                      b_x - a_x \qquad & b_y - a_y \qquad & \phantom{\lVert}b_z - a_z \\
                      c_x - a_x \qquad & c_y - a_y \qquad & \phantom{\lVert}c_z - a_z \\
                      d_x - a_x \qquad & d_y - a_y \qquad & \phantom{\lVert}d_z - a_z \\
                    \end{vmatrix}\hspace{1.57cm}&&\hspace{3cm}\nonumber
\end{flalign}

\vspace{1em}

Being completely determined by the tetrahedron vertex coordinates,
the four $3 \times 3$ determinants
can be pre-computed and stored in the tetrahedron data structure when it is created.
The cost of the {\tt inSphere} predicate becomes then negligible.
Notice also that the fourth sub-determinant 
is minus the tetrahedron volume.
We can set it to a positive value to flag deleted tetrahedra during the breadth-first search, 
thereby saving memory space.

The maximal improvement of the \textsc{Cavity}  function obtained with this optimization 
depends on the number of times the {\tt inSphere} predicate is invoked per tetrahedron.
First, in order to simplify further discussion, we will assume that the number of tetrahedra 
is about $6$ times the number of vertices in the final triangulation\cite{AOMD}.
This means that each point insertion results in average in the creation of $6$ new tetrahedra. 
On the other hand, we have seen in Section~\ref{sec:spatial_sorting} 
that an appropriate point ordering ensures 
an approximately constant number of tetrahedra in the cavities. 
This number is close to $20$
in a usual mesh generation context\cite{DBLP:journals/cad/Remacle17}.
One point insertion thus results in the deletion of $20$ tetrahedra,
and the creation of $26$ tetrahedra (all figures are approximations).
This number is also the number of triangular faces of the cavity,
since all tetrahedra created by the \textsc{DelaunayBall} function
associate a cavity facet to the inserted vertex $p_{k+1}$.
The \texttt{inSphere} predicate is therefore evaluated positively
 for the 20 tetrahedra forming the cavity and negatively for the 26 tetrahedra adjacent to the faces of the cavity,
a total of $46$ calls for each vertex insertion.

When $n$ points have been inserted in the mesh,
a total of $26 n$ tetrahedra were created 
and the predicate has been evaluated $46 n$ times. 
Thus, we may conclude from this analysis
that the \texttt{inSphere} predicate is called approximately $46n/26n = 1.77$ times per tetrahedron.
In consequence, the maximal improvement that can be obtained from our optimization of the {\tt inSphere} predicate 
is of $1 - 1/1.77 = 43\%$.
Storing the sub-determinants has a memory cost (4 {\tt double} values per tetrahedron) 
and a bandwidth cost (loading and storing of sub-determinants). 
For instance, for $n=4.\,10^6$,  
we observe a speedup of $32\%$ in the {\tt inSphere} predicate evaluations, 
taking into account the time spent to compute the stored sub-determinants.  

Note that a second geometric predicate is extensively used in Delaunay triangulation. 
The \texttt{orient3d} predicate  evaluates whether a point $d$ is 
above/on/under the plane defined by three points $\{a,b,c\} \in \reals^3$\cite{shewchuk1997adaptive}.
It computes the triple product $(b-a)\cdot((c-a)\times(d-a))$, 
i.e. the signed volume of tetrahedron $\{a,b,c,d\}$.
This predicate is mostly used in the \textsc{Walk} function. 

\paragraph{Implementation details} 

Geometrical predicates evaluated with standard floating point arithmetics 
may lead to inaccurate or inconsistent results. 
To have a robust and efficient implementations 
of the \texttt{inSphere}  and \texttt{orient3d} predicates,
we have applied the strategy implemented in Tetgen\cite{DBLP:journals/toms/Si15}.
\begin{enumerate}

	\item Evaluate first the predicate with floating point precision.
	This gives a correct value in the vast majority of the cases, 
        and represents only 1\% of the code.
	
	\item Use a filter to check whether the obtained result is certain.
	A static filter is first used, then, if more precision is needed, a dynamic filter is evaluated.
        If the result obtained by standard arithmetics is not trustworthy,
        the predicate is computed with exact arithmetics\cite{shewchuk1997adaptive}.
	
	\item To be uniquely defined, Delaunay triangulations requires each point to be 
        inside/outside a sphere, under/above a plane.
	When a point is exactly on a sphere or a plane, 
        the point is in a non-general position that is slightly perturbed 
	to ``escape'' the singularity.  
	We implemented the symbolic perturbations 
        proposed by Edelsbrunner\cite{DBLP:journals/tog/EdelsbrunnerM90}.
\end{enumerate}

\subsection{Improving {\sc DelaunayBall}: spending less time computing adjacencies} \label{sec:ball}

The {\sc DelaunayBall} function creates the tetrahedra filling the cavity (Figure~\ref{fig:delaunayball}),
and updates the tetrahedron structures. 
In particular, tetrahedron adjacencies are recomputed.
This is the most expensive step of the Delaunay triangulation algorithm 
as it typically takes about $40\%$ of the total time (Table~\ref{table:profiling_sequential}).

In contrast to existing implementations, we strongly interconnect the 
cavity building and cavity retriangulation steps.
Instead of relying on a set of elegant and independent elementary operations
like tetrahedron creation/deletion or adjacency computation,
a specific cavity data structure has been developed,
which optimizes batches of operations 
for the specific requirements of the Delaunay triangulation (Listing~\ref{cavity_struct}).

\subparagraph{Use cavity building information}
The tetrahedra reached by the breadth-first search 
during the {\sc Cavity} step (Section~\ref{sec:cavity}), 
are either inside or adjacent to the cavity. 
Each triangular facet of the cavity boundary is thus shared 
by a tetrahedron $t_1 \in {\mathcal A}({\DT}_k,{p}_{k+1})$ outside the cavity,
by a tetrahedron $t_2 \in {\mathcal C}({\DT}_k,{p}_{k+1})$ inside the cavity,
and by a newly created tetrahedron inside the cavity $t_3 \in {\mathcal B}({\DT}_k,{p}_{k+1})$ (Figure~\ref{fig:cavity_and_boundary}).
The facet of  $t_1$ adjacent to the surface of the cavity defines, 
along with the point $p_{k+1}$, the tetrahedron $t_3$.
We thus know a priori that $t_3$ is adjacent to $t_1$,
and store this information in the \texttt{cavityBoundaryFacet\_t} structure (Listing~\ref{cavity_struct}).

\subparagraph{Fast adjacencies between new tetrahedra}\label{sec:adjacencies}
During the cavity construction procedure, the list of vertices on the cavity
as well as adjacencies with the tetrahedra outside the cavity are computed. 
The last step is to compute the 
adjacencies between the new tetrahedra built inside the cavity, 
i.e. the tetrahedra of the Delaunay ball.

The first vertex of all created tetrahedra is set to be $p_{k+1}$, 
whereas the other three vertices $\{p_1,p_2,p_3\}$ are on the cavity boundary,
and are ordered so that the volume of the tetrahedral element is positive,
i.e., \texttt{orient3d($p_{k+1}, p_1, p_2, p_3)>0$}. 
As explained in the previous section,
the adjacency stored at index $4 t_i+0$, which corresponds to the facet of tetrahedron $t_i$ opposite to the vertex $p_{k+1}$,
is already known for every tetrahedron $t_i$ in ${\mathcal B}({\DT}_k,{p}_{k+1})$.
Three neighbors are thus still to be determined for each tetrahedron $t_i$,
which means practically that an adjacency index has to be attributed 
to $4 t_i+1$, $4t_i+2$ and $4t_i+3$. 
The internal facets across which these adjacencies have to be identified 
are made of the common vertex $p_{k+1}$ 
and one oriented edge of the cavity boundary.


Using an elaborated hash table with complex collision handling 
would be overkill. 
We prefer to use a double entry lookup table of dimension $n \times n$,
whose rows and columns are associated 
with the $n$ vertices $\{p_j\}$ of the cavity boundary,
to which auxiliary indices $0 \leq i_j < n$
are affected for convenience
and stored in the padding variable of the vertex,
With this, the unique index of an oriented edge $p_1p_2$ 
is set to be $n\times i_1 + i_2$,
which corresponds to one position in the $n \times n$ lookup table. 
So, for each tetrahedron $t_i$ in the Delaunay ball
with vertices $\{p_{k+1}, p_1, p_2, p_3\}$,
the adjacency index $4 t_i+1$ is stored at position $n \times i_2 + i_3$ in the lookup table,
and similarly 
$4 t_i+2$ at position $n \times i_3 + i_1$ and
$4 t_i+3$ at position $n \times i_1 + i_2$.
A square array with a zero diagonal is built proceeding this way,
in which the sought adjacencies are the pairs of symmetric components. 

Theoretically, there is no maximal value for the number $n$ of vertices in the cavity
but, in practice, we can take advantage of the fact that it remains relatively small.
Indeed, the cavity boundary is the triangulation of a topological sphere, i.e. 
a planar graph in which the number of edges is $E= 3F/2$, where $F$ is the number of faces, 
and whose Euler characteristic is $n-E+F=2$.
Hence, the number of vertices is linked to the number of triangular faces by $n = F/2 +2$. 
As we have shown earlier that $F$ is about 26,
we deduce that there are about $n=15$ vertices on a cavity boundary. 
Hence, a $n_{max} \times n_{max}$ lookup table, with maximum size $n_{max} = 32$
is sufficient provided there are  
at most $F_{max} = 2(n_{max}-2)= 60$ tetrahedra created in the cavity.
If the cavity exceptionally contains more elements, 
the algorithm smoothly switches to a simple linear search.

Once all adjacencies of the new tetrahedra have been properly identified, 
the \textsc{DelaunayBall} function is then in charge of updating the mesh data structure.
This ends the insertion of point $p_{k+1}$.
The space freed by deleted tetrahedra is reused, if needed additional tetrahedra are added.
Note that almost all steps of adjacencies recovery are vectorizable. 

%
%

%

\begin{lstlisting}[caption={Cavity specific data structure.}, label=cavity_struct, captionpos=b, float]
typedef struct {
    uint32_t new_tetrahedron_vertices[4];  //  facet vertices + vertex to insert
    uint64_t adjacent_tetrahedron_ID; 
} cavityBoundaryFacet_t

typedef struct{
    uint64_t adjacency_map[1024];  // optimization purposes, see Section *@\ref{sec:adjacencies}@*

    struct {
        cavityBoundaryFacet_t* boundary_facets;
        uint64_t num;           // number of boundary facets
        uint64_t allocated_num; // capacity [in cavityBoundaryFacet_t]
    } to_create;

    struct {
        uint64_t* tetrahedra_ID;
        uint64_t num;           // number of deleted tetrahedra
        uint64_t allocated_num; // capacity
    } deleted;
} cavity_t;
\end{lstlisting}

\begin{figure}
  \begin{minipage}[c]{0.4\textwidth}
 		\centering
        \vspace{-0.1cm}
		\includegraphics[width=\textwidth]{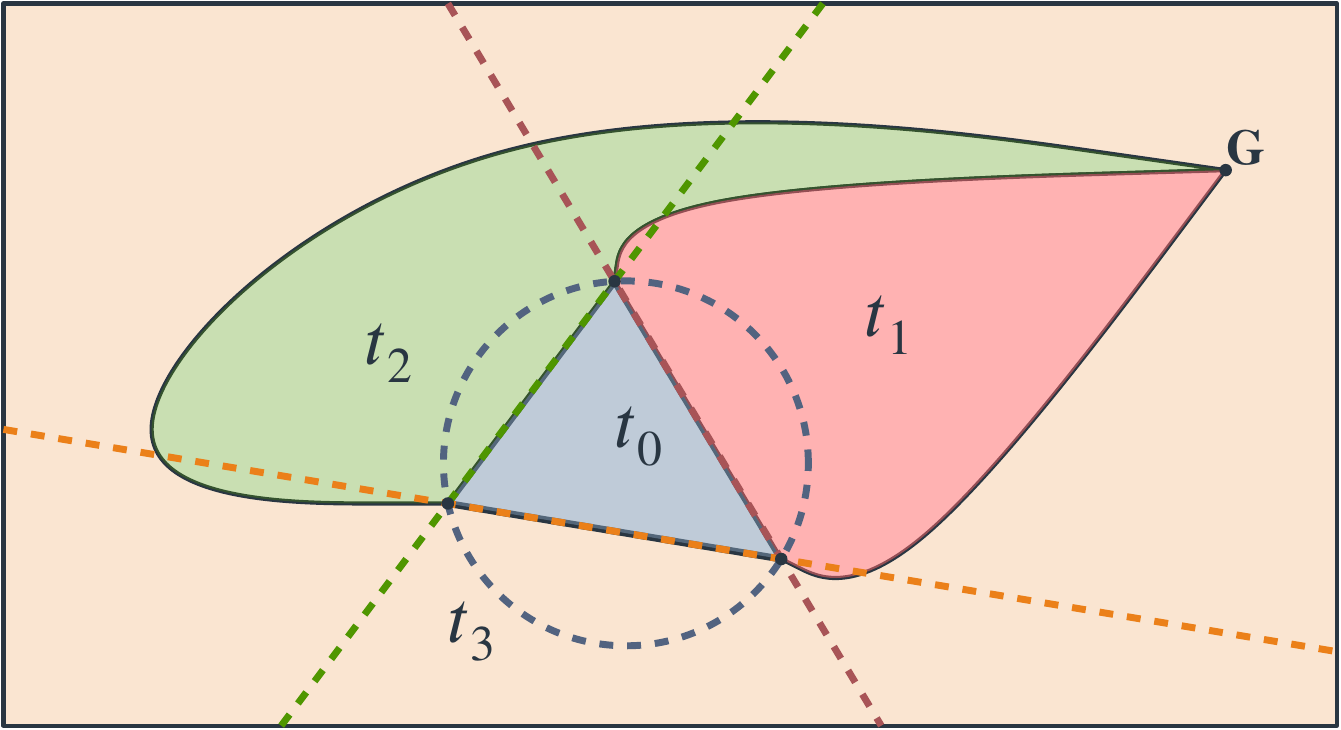}
		\caption{Triangle $t_0$ surrounded by three ghost triangles $t_1$, $t_2$ and $t_3$. Circumcircles are shown in dash lines of the respective color.}
		\label{fig:ghost_2D}
	\end{minipage}\hfill
   \begin{minipage}[c]{0.4\textwidth}
   		\centering
        \vspace{-0.2cm}
		\includegraphics[width=0.663\textwidth]{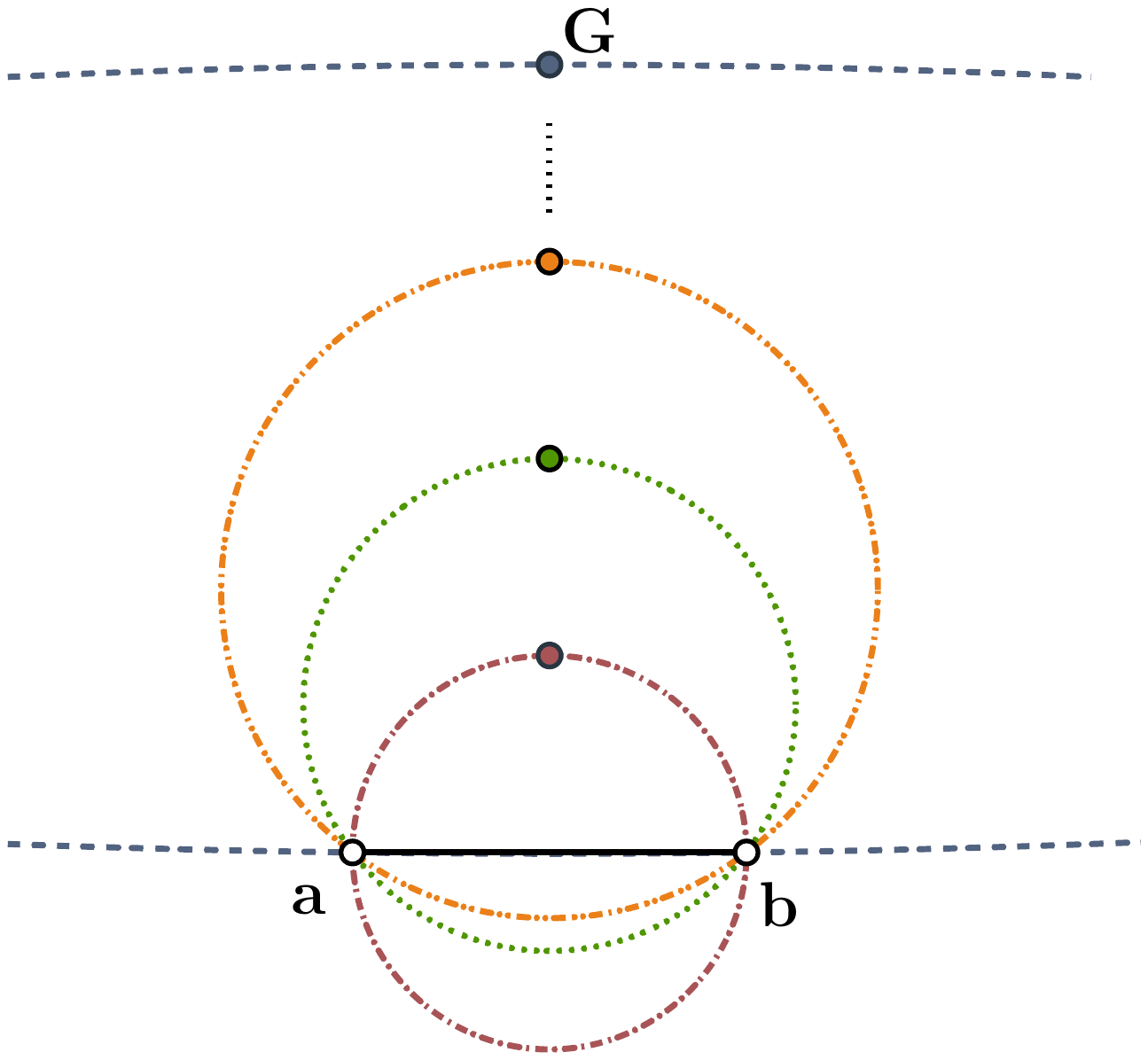}
		\caption{Circumcircles of an edge $\overline{ab}$ and increasingly distant vertices. For a vertex $G$ infinitely far away, the degenerate circle approaches a line.}
		\label{fig:ghost_circle}
  \end{minipage}
\end{figure}

\subsection{About a ghost}\label{sec:ghost}


The Bowyer–Watson algorithm for Delaunay triangulation assumes that
all newly inserted vertices 
are inside an element of the triangulation at the previous step (Algorithm~\ref{algo:seqDelaunay}).
To insert a point $p_{k+1}$ outside the current support of the triangulation,
one possible strategy is to enclose the input vertices in a sufficiently large bounding box,
and to remove the tetrahedra lying outside the convex hull of the point set 
at the end of the algorithm.
A more efficient strategy, adopted in TetGen\cite{DBLP:journals/toms/Si15}, CGAL
\cite{boissonnat2000triangulations}, Geogram \cite{levy2015geogram},
is to work with the so-called ghost tetrahedra 
connecting the exterior faces of the triangulation with a virtual ghost vertex $G$.
Using this elegant concept, 
vertices can be inserted outside the current triangulation support 
with almost no algorithmic change.

The ghost vertex $G$ is the vertex "at infinity" 
shared by all ghost tetrahedra (Figure~\ref{fig:ghost_2D}).
The ghost tetrahedra cover the whole space outside the regular triangulation. 
Like regular tetrahedra, ghost tetrahedra are stored in the mesh data structure, 
and are deleted whenever a vertex is inserted inside their circumsphere.
The accurate definition of the circumsphere of a ghost tetrahedron,
in particular with respect to the requirements of the Delaunay condition,
is however a more delicate question.

\begin{figure}[b]
\begin{subfigure}[b]{\textwidth}
\begin{subfigure}[b]{0.56\textwidth}
    \setlength\figureheight{0.6\textwidth}
    \setlength\figurewidth{\textwidth}
    \begin{tikzpicture}

\begin{axis}[%
width=\figurewidth,
height=\figureheight,
xmode=log,
xmin=10000,
xmax=10000000,
xlabel={Number of points (random uniform distribution)},
ymode=log,
ymin=0.02,
ymax=100,
ylabel={Time [s]},
log y ticks with fixed point,
]

\addplot
  table[row sep=crcr]{%
0000000010 0.000010 \\
0000000020 0.000025 \\
0000000050 0.000084 \\
0000000100 0.000182 \\
0000000200 0.000501 \\
0000000500 0.001449 \\
0000001000 0.003226 \\
0000002000 0.007427 \\
0000005000 0.016679 \\
0000010000 0.027191 \\
0000020000 0.047044 \\
0000050000 0.113428 \\
0000100000 0.214826 \\
0000200000 0.406734 \\
0000500000 1.072637 \\
0001000000 2.034790 \\
0002000000 4.288870 \\
0005000000 10.558001 \\
0010000000 21.655637 \\
};
\addlegendentry{Ours};



\addplot
  table[row sep=crcr]{%
0000000010 0.010000 \\
0000000020 0.010000 \\
0000000050 0.010000 \\
0000000100 0.010000 \\
0000000200 0.010000 \\
0000000500 0.010000 \\
0000001000 0.010000 \\
0000002000 0.020000 \\
0000005000 0.030000 \\
0000010000 0.060000 \\
0000020000 0.110000 \\
0000050000 0.260000 \\
0000100000 0.510000 \\
0000200000 1.030000 \\
0000500000 2.635000 \\
0001000000 5.525000 \\
0002000000 10.990000 \\
0005000000 28.160000 \\
0010000000 56.015000 \\
};
\addlegendentry{Geogram 1.5.4};

\addplot
  table[row sep=crcr]{%
0000000010 0.000039 \\
0000000020 0.000081 \\
0000000050 0.000234 \\
0000000100 0.000552 \\
0000000200 0.001228 \\
0000000500 0.003486 \\
0000001000 0.006949 \\
0000002000 0.012404 \\
0000005000 0.030542 \\
0000010000 0.062383 \\
0000020000 0.124289 \\
0000050000 0.321103 \\
0000100000 0.643358 \\
0000200000 1.297371 \\
0000500000 3.343558 \\
0001000000 6.645409 \\
0002000000 13.199438 \\
0005000000 33.081317 \\
0010000000 66.244400 \\
};
\addlegendentry{CGAL 4.12};

\addplot
  table[row sep=crcr]{%
0000000010 0.000038 \\
0000000020 0.000078 \\
0000000050 0.000217 \\
0000000100 0.000464 \\
0000000200 0.001062 \\
0000000500 0.002927 \\
0000001000 0.006219 \\
0000002000 0.011761 \\
0000005000 0.026850 \\
0000010000 0.054123 \\
0000020000 0.109306 \\
0000050000 0.284304 \\
0000100000 0.564524 \\
0000200000 1.161660 \\
0000500000 2.924567 \\
0001000000 5.892280 \\
0002000000 12.209800 \\
0005000000 31.314933 \\
0010000000 63.994833 \\
};
\addlegendentry{TetGen 1.5.1-beta1};

\end{axis}
\end{tikzpicture}%
\end{subfigure}
~
\begin{subfigure}[b]{0.38\textwidth}
    \begin{tabular}{L{1.4cm} r r r r}
	\toprule
    \# vertices & $10^4$ & $10^5$ & $10^6$ & $10^7$ \\
	\midrule
    \textcolor{UCLdarkBlue!80!black}{Ours} & 0.027 & 0.21 & 2.03 & 21.66 \\
    \textcolor{UCLorange!80!black}{Geogram} & 0.060 & 0.51 & 5.53 & 56.02\\
    \textcolor{UCLgreen!80!black}{CGAL} & 0.062 & 0.64 & 6.65 & 66.24\\
    \textcolor{UCLdarkRed!80!black}{TetGen} & 0.054 & 0.56 & 5.89 & 63.99\\
	\bottomrule
    \end{tabular}
    \vspace{1.5cm}
\end{subfigure}
\caption{Intel$^\circledR$ Core$^\text{TM}$ i7-6700HQ CPU, maximum core frequency of $3.5$Ghz.}
\label{fig:sequential_i7}
\end{subfigure}

\vspace{0.25cm}

\begin{subfigure}[b]{\textwidth}
\begin{subfigure}[b]{0.56\textwidth}
    \setlength\figureheight{0.6\textwidth}
    \setlength\figurewidth{\textwidth}
    \begin{tikzpicture}

\begin{axis}[
width=\figurewidth,
height=\figureheight,
xmode=log,
xmin=10000,
xmax=50000000,
xlabel={Number of points (random uniform distribution)},
ymode=log,
ymin=0.1,
ymax=10000,
ylabel={Time [s]},
log y ticks with fixed point,
]

\addplot
  table[row sep=crcr]{
0000000010 0.000066 \\
0000000020 0.000115 \\
0000000050 0.000334 \\
0000000100 0.000756 \\
0000000200 0.001584 \\
0000000500 0.004664 \\
0000001000 0.010010 \\
0000002000 0.024177 \\
0000005000 0.085996 \\
0000010000 0.133879 \\
0000020000 0.224012 \\
0000050000 0.523731 \\
0000100000 0.982567 \\
0000200000 1.891721 \\
0000500000 4.864827 \\
0001000000 9.357716 \\
0002000000 19.290039 \\
0005000000 48.406428 \\
0010000000 97.972996 \\
0020000000 228.630455 \\
0050000000 624.790906 \\
};
\addlegendentry{Ours};


\addplot
  table[row sep=crcr]{
0000000010 0.010000 \\
0000000020 0.010000 \\
0000000050 0.010000 \\
0000000100 0.010000 \\
0000000200 0.010000 \\
0000000500 0.010000 \\
0000001000 0.030000 \\
0000002000 0.050000 \\
0000005000 0.120000 \\
0000010000 0.240000 \\
0000020000 0.480000 \\
0000050000 1.220000 \\
0000100000 2.470000 \\
0000200000 4.970000 \\
0000500000 12.580000 \\
0001000000 25.340000 \\
0002000000 50.860000 \\
0005000000 129.325000 \\
0010000000 259.740000 \\
0020000000 522.716667 \\
0050000000 1354.916667 \\
};
\addlegendentry{Geogram 1.5.4};

\addplot
  table[row sep=crcr]{
0000000010 0.000248 \\
0000000020 0.000387 \\
0000000050 0.000943 \\
0000000100 0.002053 \\
0000000200 0.004347 \\
0000000500 0.011304 \\
0000001000 0.023873 \\
0000002000 0.049679 \\
0000005000 0.128386 \\
0000010000 0.264736 \\
0000020000 0.545270 \\
0000050000 1.388924 \\
0000100000 2.808278 \\
0000200000 5.698767 \\
0000500000 14.330397 \\
0001000000 28.362427 \\
0002000000 57.115567 \\
0005000000 143.332246 \\
0010000000 286.544046 \\
0020000000 572.836334 \\
0050000000 1440.008377 \\
};
\addlegendentry{CGAL 4.12};

\addplot
  table[row sep=crcr]{
0000000010 0.001 \\
0000000020 0.001 \\
0000000050 0.001 \\
0000000100 0.002 \\
0000000200 0.004 \\
0000000500 0.010000 \\
0000001000 0.020000 \\
0000002000 0.050000 \\
0000005000 0.136667 \\
0000010000 0.283333 \\
0000020000 0.570000 \\
0000050000 1.456667 \\
0000100000 2.970000 \\
0000200000 6.040000 \\
0000500000 15.433333 \\
0001000000 31.130000 \\
0002000000 63.526667 \\
0005000000 162.906667 \\
0010000000 336.206667 \\
0020000000 704.823333 \\
0050000000 1918.720000 \\
};
\addlegendentry{TetGen 1.5.1-beta1};

\end{axis}
\end{tikzpicture}
\end{subfigure}
~
\begin{subfigure}[b]{0.38\textwidth}
    \begin{tabular}{L{1.4cm} r r r r}
	\toprule
    \# vertices & $10^4$ & $10^5$ & $10^6$ & $10^7$ \\
	\midrule
    \textcolor{UCLdarkBlue!80!black}{Ours} & 0.134 & 0.98 & 9.36 & 97.97\\
    \textcolor{UCLorange!80!black}{Geogram} & 0.240 & 2.47 & 25.34 & 259.74\\
    \textcolor{UCLgreen!80!black}{CGAL} & 0.265 & 2.81 & 28.36 & 286.54\\
    \textcolor{UCLdarkRed!80!black}{TetGen} & 0.283 & 2.97 & 31.13 & 336.21\\
	\bottomrule
    \end{tabular}
    \vspace{1.5cm}
\end{subfigure}
    \caption{Intel$^\circledR$ Xeon Phi$^\text{TM}$ 7210 CPU, maximum core frequency of $1.5$Ghz.}
    \label{fig:sequential_KNL}
\end{subfigure}
	 \caption{Performances of our sequential Delaunay triangulation implementation (Algorithm~\ref{algo:seqDelaunay})
	 on a laptop \textbf{(a)} and on a slow CPU having AVX-512 vectorized instructions \textbf{(b)}. Timings are in seconds and exclude the initial spatial sort.}
\end{figure}

From a geometrical point of view,
as the ghost vertex $G$ moves away towards infinity 
from an exterior facet $abc$ of the triangulation,
the circumscribed sphere of tetrahedron $abcG$ tends to the half space
on the positive side of the plane determined by the points $abc$,
i.e., the set of points $x$ such that \texttt{orient3d($a,b,c,d$)} is positive. 
The question is whether this inequality should be strict or not,
and the robust answer is neither one nor the other.
This is illustrated in 2D in Figure~\ref{fig:ghost_circle}.
The circumcircle of a ghost triangle $abG$ actually contains 
not only the open half-plane strictly above the line $\overleftrightarrow{ab}$,
but also the line segment $[ab]$ itself;
the other parts of the line $\overleftrightarrow{ab}$ being excluded. 
In 3D, the circumsphere of a ghost tetrahedron $abcG$ 
contains the half-space $L$ such that for any point $d \in L$, 
\texttt{orient3d($a,b,c,d$)} is strictly positive,
plus the disk defined by the circumcircle of the triangle $abc$. 
If the point $d$ is in the plane $abc$, 
i.e. \texttt{orient3d($a,b,c,d$)=0}, 
we thus additionally test if $d$ is in the circumsphere
of the adjacent regular tetrahedron sharing the triangle $abc$. 
This composite definition of the circumscribed circle
makes this approach robust with minimal changes:
only the \texttt{inSphere} predicate is modified in the implementation.

Because the $3 \times 3$ determinant of \texttt{orient3d} is used
instead of the $4 \times4 $ determinant of \texttt{inSphere} for ghost tetrahedra, 
the approach with a ghost vertex is faster than other strategies \cite{shewchuk1997delaunay}.
Note that the {\sc Walk} cannot evaluate \texttt{orient3d} on the faces of ghost tetrahedra that connect edges of the convex hull to the ghost vertex.
If the starting tetrahedron is a ghost tetrahedron, the {\sc Walk} directly steps into the non-ghost adjacent tetrahedron.
Then, if it steps in an other ghost tetrahedron $abcG$ while walking toward $p_{k+1}$, 
we have \texttt{orient3d($a,b,c,p_{k+1})>0$} and the ghost tetrahedron is inside of the cavity. The walk hence stops there, and the algorithm proceeds as usual.

%
\subsection{Serial implementation performances}

Our reference implementation of the serial Delaunay triangulation algorithm has about one thousand lines
(without Shewchuk's geometric predicates\cite{shewchuk1997adaptive}).
It is open-source and available in  Gmsh (\url{www.gmsh.info}). 
Overall, it is almost three time faster than other sequential implementations. 
Figure \ref{fig:sequential_i7} and \ref{fig:sequential_KNL} 
show the performance gap between our implementation and concurrent software 
on a laptop with a maximum core frequency of $3.5$Ghz,
and on a many-core computer with a maximum core frequency of $1.5$Ghz and wider SIMD instructions.
Table~\ref{table:profiling_sequential} indicates that the main difference in speed 
comes from the more efficient adjacencies computation in the \texttt{DelaunayBall} function. 
Other software use a more general function 
that can also handle cavities with multiple interior vertices. 
But this situation does not happen with Delaunay insertion, 
where there is always one unique point in the cavity, $p_{k+1}$.
Since our \texttt{DelaunayBall} function is optimized for Delaunay cavities, 
it is approximately three time faster,
despite  the additional computation of sub-determinants.
The remaining performance gain 
is explained by our choice of simple but efficient memory aligned data structure.

\section{Parallel Delaunay}\label{sec:parallel}

\subsection{Related work}  

To overcome memory and time limitations, 
Delaunay triangulations should be constructed in parallel making the most 
of both distributed and shared memory architectures.
A triangulation can be subdivided into multiple parts,
each constructed and stored on a node of a distributed memory cluster.
Multiple methods have been proposed to merge independently generated Delaunay triangulations\cite{cignoni_parallel_1993, chen_merge_2010, funke_parallel_2017, blelloch_design_1999}.
However, those methods feature complicated merge steps,
which are often difficult to parallelize.
To avoid merge operations, other distributed implementations maintain a single valid Delaunay triangulation and use synchronization between processors whenever a conflict may occur at inter-processor boundaries\cite{okusanya_3d_nodate, chrisochoides_parallel_2003}.
In finite-element mesh generation, merging two triangulations can be simpler because triangulations are not required to be fully Delaunay, allowing algorithms to focus primarily on load balancing\cite{lachat_parallel_nodate}.

On shared memory machines, divide-and-conquer approaches remain efficient, but other approaches have been proposed since communication costs between different threads are not prohibitive 
to the contrary of distributed memory machines.
To insert a point in a Delaunay triangulation,
the kernel procedure operates on a cavity that 
is modified to accommodate the inserted point (Figure~\ref{fig:cavity_and_boundary}).
Two points can therefore be inserted concurrently in a Delaunay triangulation 
if their respective cavities do not intersect, 
$ {\mathcal C} (DT_k, p_{k1}) \cap {\mathcal C}(DT_k, p_{k2}) = \emptyset$,
otherwise there is a conflict.
In practice, other types of conflicts and data-races should possibly be taken into account
depending on the chosen data structures and the insertion point strategy.
Conflict management strategies relying heavily on locks
\footnote{A lock is a synchronization mechanism enforcing that multiple threads do not access a resource at the same time. When a thread cannot acquire a lock, it usually waits.}
lead to relatively good speedups on small numbers of cores\cite{kohout_parallel_2005,blandford_engineering_2006,batista_parallel_2010,chriso_2012}. Remacle et al. presented an interesting strategy that checks if insertions can be done in parallel by synchronizing threads with barriers\cite{DBLP:journals/cad/Remacle17}. However, synchronization overheads prevent those strategies from scaling to an high number of cores. More promising approaches rely on a partitioning strategy\cite{lo_parallel_2012, LOSEILLE201557}. Contrarily to pure divide-and-conquer strategies for distributed memory machines, partitions can change and move regularly, and they do not need to cover the whole mesh at once.

In this section, we propose to parallelize the Delaunay kernel using partitions based on a space-filling curve, similarly to Loseille et al.\cite{LOSEILLE201557}. The main difference is that we significantly modify the partitions at each iteration level. Our code is designed for shared memory architecture only, we leave its integration into a distributed implementation for future work.

\subsection{A parallel strategy based on partitions}

There are multiple conditions a program should ensure 
to avoid data-races and conflicts between threads
when concurrently inserting points in a Delaunay triangulation.
Consider thread $t_1$ is inserting point $p_{k1}$ and thread $t_2$ is simultaneously inserting point $p_{k2}$.

\begin{enumerate}[ref=(\arabic*)]
    \item Thread $t_1$ cannot access information about any tetrahedron in ${\mathcal C}({\DT},p_{k2})$ and inversely. Hence: \label{enum:rule1}
    \begin{enumerate}[ref=(\arabic{enumi}\alph*)]
        \item ${\mathcal C}({\DT},p_{k1}) \cap {\mathcal C}({\DT},p_{k2})  = \emptyset$ \label{enum:rule1a}
        \item ${\mathcal A}({\DT},p_{k1}) \cap {\mathcal C}({\DT},p_{k2}) = \emptyset$ and 
              ${\mathcal A}({\DT},p_{k2}) \cap {\mathcal C}({\DT},p_{k1}) = \emptyset$\label{enum:rule1b}
        \item Thread $t_1$ cannot walk into ${\mathcal C}({\DT},p_{k2})$ and reciprocally,
                     $t_2$ cannot walk into ${\mathcal C}({\DT},p_{k1})$ \label{enum:rule1c}
    \end{enumerate}
    \item A tetrahedron in ${\mathcal B}({\DT},p_{k1})$ and a tetrahedron in ${\mathcal B}({\DT},p_{k2})$ cannot be created at the same memory index. \label{enum:rule2}
\end{enumerate}

\begin{figure}[b]
    \centering
    \includegraphics[width=0.7\textwidth]{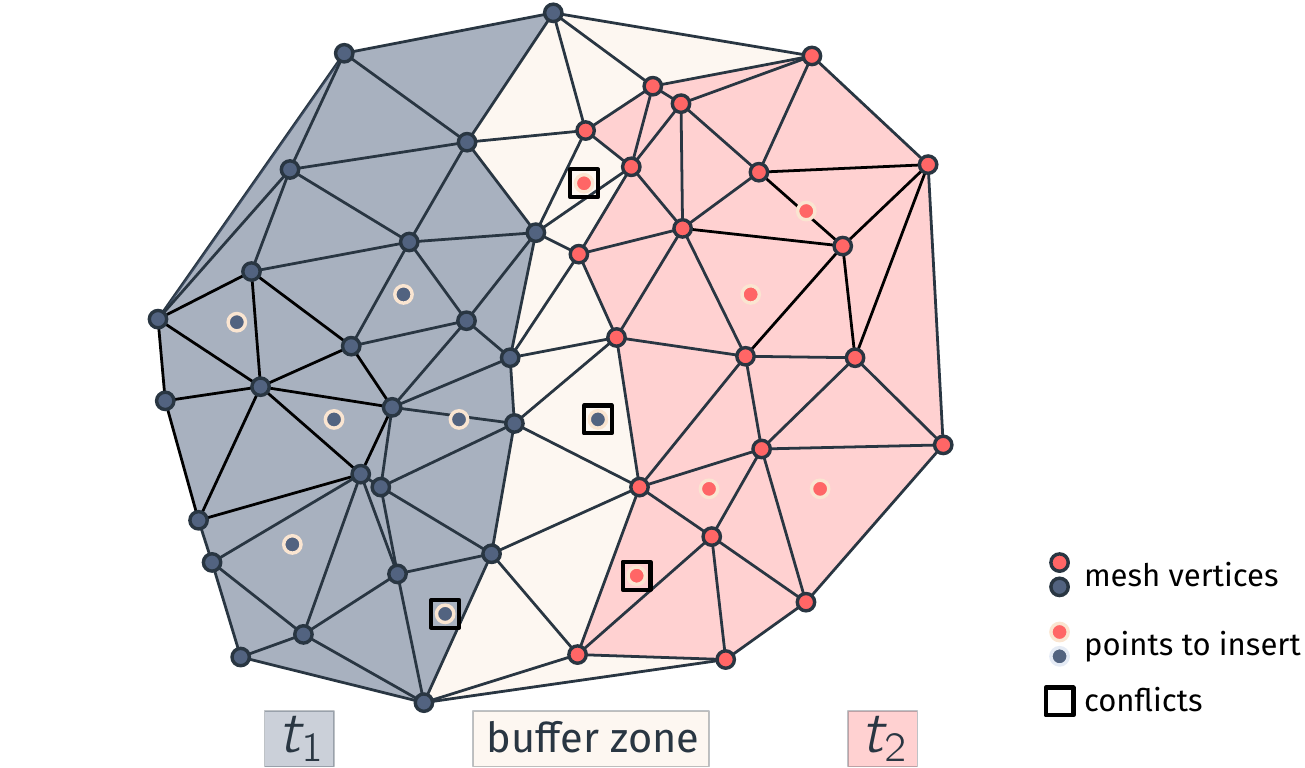}
    \caption{Vertices are partitioned such that each vertex belongs to a single thread. A triangle can only be modified by a thread that owns all of its three vertices. Triangles that cannot be modified by any thread form a buffer zone.}
    \label{fig:conflict}
\end{figure}

To ensure rule \ref*{enum:rule1} it is sufficient to restrain each thread to work on an independent partition
of the mesh (Figure \ref{fig:conflict}).
This lock-free strategy minimizes synchronization between threads and is very efficient.
Each point of the Delaunay triangulation is assigned a partition corresponding to a unique thread.
A tetrahedron belongs to a thread if at least three of its vertices are in that thread's partition.

To ensure \ref*{enum:rule1a} and \ref*{enum:rule1b},
the insertion of a point belonging to thread 
is aborted if the thread accessed a tetrahedron that belongs to another thread or that is in the buffer zone.
To ensure \ref*{enum:rule1c}, we forbid threads to walk in tetrahedra belonging to another thread.
Consequently, a thread aborts the insertion of a vertex when 
(i) the \textsc{Walk} reaches another partition, or when 
(ii) the \textsc{Cavity} function reaches a tetrahedron in the buffer zone.
To insert these points, the vertices are re-partitioned differently (Section \ref{sec:partitioning}), 
a procedure repeated until there is no more vertices to insert or until 
the rate of successful insertions has become too low.
In that case, the number of threads is decreased (Section \ref{subsec:reduction}).
When the number of points to insert has become small enough, 
the insertion runs in sequential mode to insert all remaining vertices.
The first BRIO round is also inserted sequentially to generate a first base mesh.
Nevertheless, the vast majority of points are inserted in parallel.

Rule \ref*{enum:rule2} is obvious from a parallel memory management point of view. 
It is however difficult to ensure it without requiring an unreasonable amount of memory. 
As explained in Section \ref{subsec:critical}, synchronization between threads is required punctually.

\subsection{Partitioning and re-partitioning with Moore curve}\label{sec:partitioning}

\begin{figure}[htb]
    \centering
    \begin{subfigure}[t]{0.42\textwidth}
        \centering
        \includegraphics[width=0.9\textwidth]{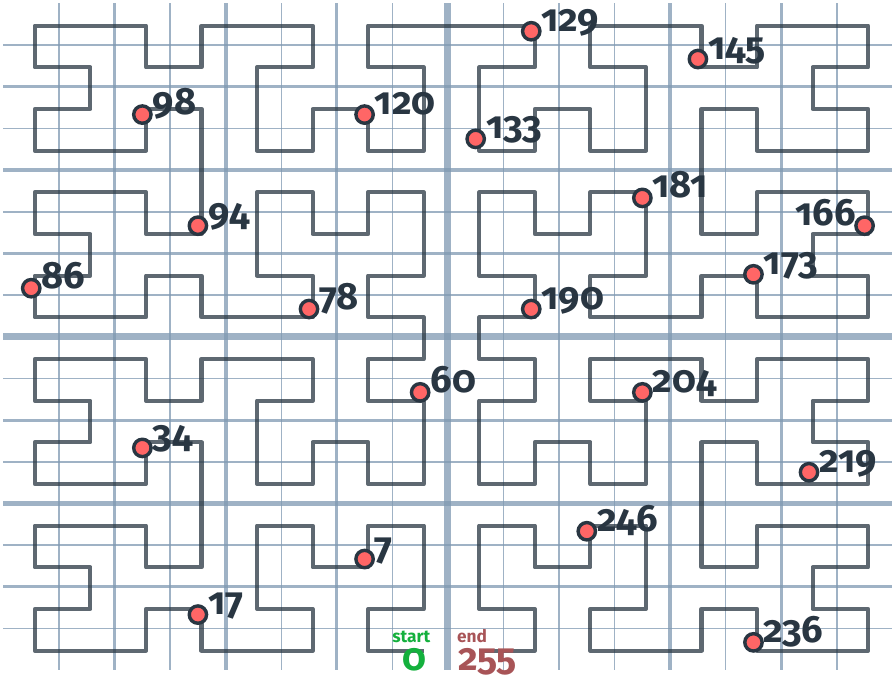}
        \caption{}
        \label{moore_a}
    \end{subfigure}
    ~~~~
    \begin{subfigure}[t]{0.42\textwidth}
        \centering
        \includegraphics[width=0.9\textwidth]{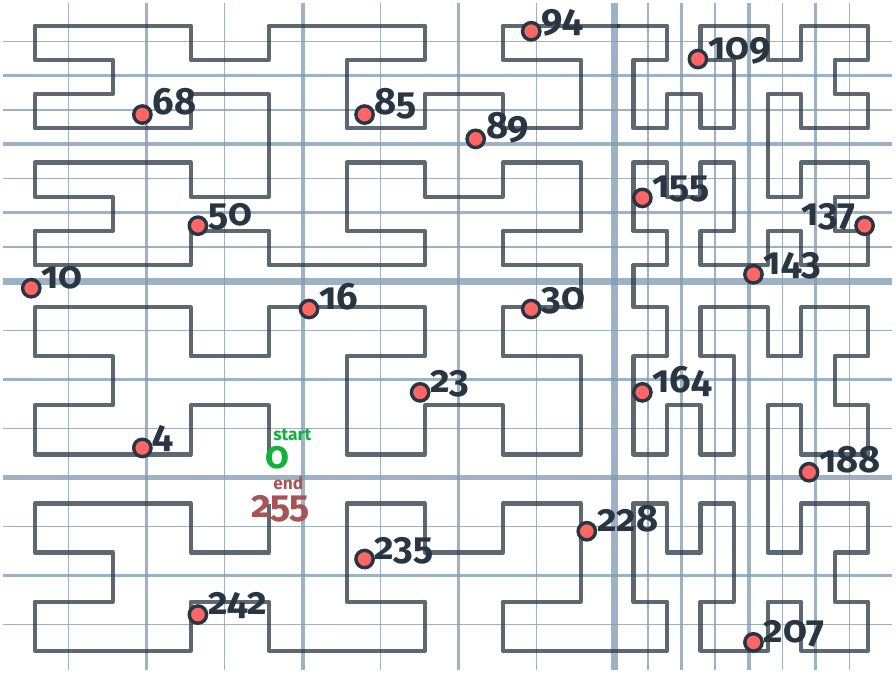}
        \caption{}
        \label{moore_b}
    \end{subfigure}
    \\[0.4cm]
    \begin{subfigure}[t]{0.42\textwidth}
        \centering
        \includegraphics[width=0.9\textwidth]{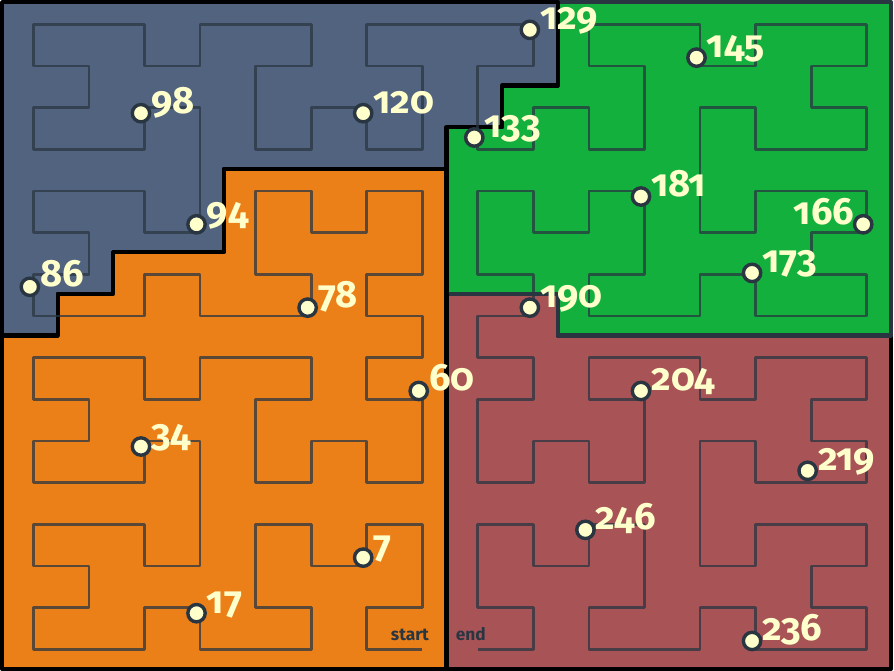}
        \begin{tabular}{l r r c}
            & & & \\[-0.2cm]
			\toprule
             & \multicolumn{2}{c}{partition} & \multirow{2}{1.5cm}{\#points} \\
             & start & end & \\\midrule
            \textcolor{UCLorange}{\bf $\mathbf{1^{\text{st}}}$ thread} & 0 & 94 & 5 \\
            \textcolor{UCLdarkBlue}{\bf $\mathbf{2^{\text{nd}}}$ thread} & 94 & 133 & 5 \\
            \textcolor{UCLgreen}{\bf $\mathbf{3^{\text{rd}}}$ thread} & 133 & 190 & 5 \\
            \textcolor{UCLdarkRed}{\bf $\mathbf{4^{\text{th}}}$ thread} & 190 & $\infty$ & 5 \\
			\bottomrule
        \end{tabular}
        \caption{}
        \label{moore_c}
    \end{subfigure}
    ~~~~
   \begin{subfigure}[t]{0.42\textwidth}
        \centering
        \includegraphics[width=0.9\textwidth]{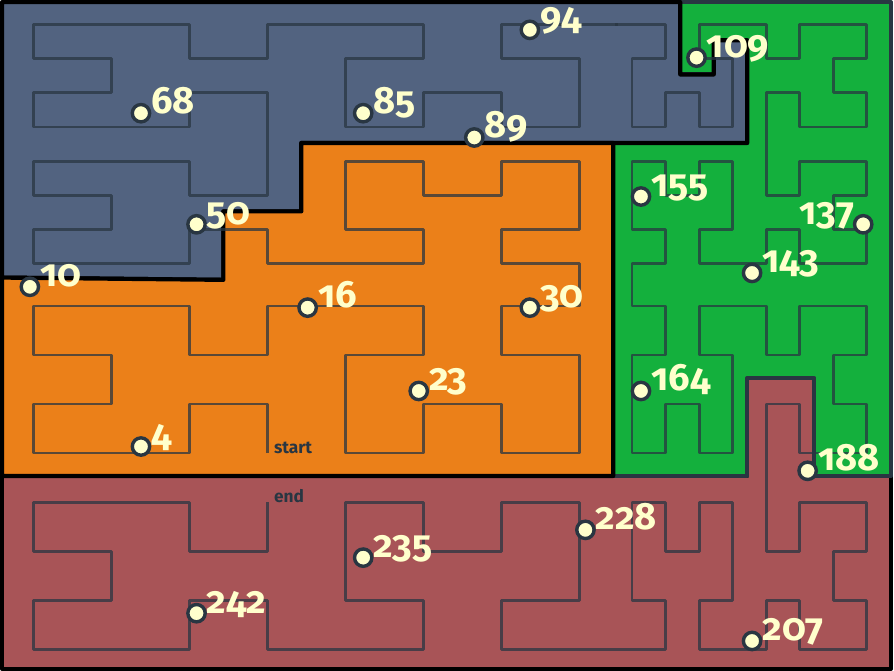}
        \begin{tabular}{l r r c}
            & & & \\[-0.2cm]
			\toprule
             & \multicolumn{2}{c}{partition} & \multirow{2}{1.5cm}{\#points} \\
             & start & end & \\\midrule
            \textcolor{UCLorange}{\bf $\mathbf{1^{\text{st}}}$ thread} & 0 & 50 & 5\\
            \textcolor{UCLdarkBlue}{\bf $\mathbf{2^{\text{nd}}}$ thread} & 50 & 109 & 5 \\
            \textcolor{UCLgreen}{\bf $\mathbf{3^{\text{rd}}}$ thread} & 109 & 188 & 5 \\
            \textcolor{UCLdarkRed}{\bf $\mathbf{4^{\text{th}}}$ thread} & 188 & $\infty$ & 5 \\
			\bottomrule
        \end{tabular}
        \caption{}
        \label{moore_d}
    \end{subfigure}
    \caption{Partitioning of 20 points in 2D using the Moore indices, on the right the supporting grid of the Moore curve is transformed and the curve is shifted. In both cases, each partition contains 5 points. Indeed, the starting and ending Moore index of each partition are defined in a way that balances the point insertions between threads. 
	}
    \label{fig:moore}
\end{figure}

We subdivide the $n_v$ points to insert such that each thread
inserts the same number of points.
Our partitioning method is based on the Moore curve, i.e. on the point insertion order implemented 
for the sequential Delaunay triangulation (Section~\ref{sec:spatial_sorting}).
Space-filling curves are relatively fast to construct and have already been used successfully for partitioning meshes \cite{fd4,634498,devine2005new,LOSEILLE201557}.
Each partition of the $n_{thread}$ partitions is a set of grid cells that are consecutive along the Moore curve. 
To compute the partitions, i.e. its starting and ending Moore indices, we sort the points to insert
according to their Moore indices (see Section~\ref{sec:spatial_sorting}).
Then, we assign the first $n_v/n_{threads}$ points to the first partition, 
the next $n_v/n_{threads}$ to the second, etc (Figure~\ref{moore_c}).
The  second step is to partition the current Delaunay triangulation in which the points will
be inserted. 
We use once more the Moore indices to assign the mesh vertices to the different partitions. The ghost vertex is assigned a random index.
The partition owning a tetrahedron is determined from the partitions of its vertices, if a tetrahedron has at least three vertices 
in a partition it belong to this partition, otherwise the tetrahedron is in the buffer zone.
Each thread then owns a subset of the points to insert and a subset of the current triangulation.

Once all threads attempted to insert all their points, a large majority of vertices is generally inserted.
To insert the vertices for which insertion failed because the point cavity 
spans multiple partitions (Figure~\ref{fig:conflict}), we modify significantly the partitions by 
modifying the Moore indices computation. We apply a coordinate transformation and 
a circular shift to move the zero index around the looping curve (Figure~\ref{fig:moore}).
Coordinates below a random threshold are linearly compressed, while coordinates above the threshold are linearly expanded.

\begin{figure}
\begin{minipage}[c]{0.48\textwidth}
    \centering
    \setlength\figureheight{0.51\textwidth}
    \setlength\figurewidth{0.88\textwidth}
    \begin{tikzpicture}
\begin{axis}[
width=0.955\figurewidth,
height=\figureheight,
at={(0,0)},
scale only axis,
xmode=log,
xmin=1,
xmax=64,
xtick={1,2,4,8,16,32,64,128,256},
xlabel={Number of threads},
ymode=log,
ymin=1,
ymax=320,
log x ticks with fixed point,
log y ticks with fixed point,
ylabel={Time [s]},
legend style={at={(0.985,0.985)}, anchor=north east, legend cell align=left,align=left, draw=UCLblack},
]

\addplot+[color=UCLorange,solid,thick, mark=*]
  table[row sep=crcr]{
1 42.528518 \\
2 25.140887 \\
4 14.439180 \\
8 12.581156 \\
};
\addlegendentry{Intel i7-6700HQ};

\addplot+[color=UCLdarkBlue,solid, thick, mark=*]
  table[row sep=crcr]{
1 181.593807 \\
2 99.922172 \\
4 51.594421 \\
8 27.763146 \\
16 15.099379 \\
32 8.899023 \\
64 5.907685 \\
128 5.793321 \\
256 7.588546 \\
};
\addlegendentry{Intel Xeon Phi 7210};

\addplot+[color=UCLdarkRed, solid, thick, mark=*]
	table[row sep=crcr]{
1 45.1856 \\
2 24.7224 \\
4 15.6832 \\
8 8.4806 \\
16 4.8806 \\
32 3.178 \\
64 2.7296 \\
128 2.6964 \\
};
\addlegendentry{AMD EPYC 7551};

\addplot+[color=UCLdarkGrey,dashed, mark=none]
  table[row sep=crcr]{
1	181.593807 \\
256 0.70935080859375 \\
};
\addlegendentry{perfect scaling};

\addplot+[color=UCLdarkGrey,dashed, mark=none]
  table[row sep=crcr]{
1	42.528518 \\
256 0.1661270234375 \\
};

\addplot+[color=UCLdarkGrey,dashed, mark=none]
table[row sep=crcr]{
1 45.1856 \\
128 0.3530125 \\
};

\end{axis}
\end{tikzpicture}
    \caption{Strong scaling of our parallel Delaunay for a random uniform distribution of 15 million points, resulting in over 100 million tetrahedra on 3 machines: a quad-core laptop, an Intel Xeon Phi with 64 cores and a dual-socket AMD EPYC $2\times 32$ cores.}
    \label{fig:hxt_parallel_scaling}
\end{minipage}\hfill
\begin{minipage}[c]{0.48\textwidth}
    \vspace{0.2cm}
    \centering
    \setlength\figureheight{0.5\textwidth}
    \setlength\figurewidth{0.88\textwidth}
	\begin{tikzpicture}

\begin{axis}[
width=0.955\figurewidth,
height=\figureheight,
at={(0,0)},
scale only axis,
xmode=log,
xmin=10000,
xmax=1000000000,
xlabel={Number of points (random uniform distribution)},
ymode=log,
ymin=0.5,
ymax=100,
ylabel={Million tetrahedra per second},
log y ticks with fixed point,
legend style={at={(0.015,0.985)}, anchor=north west, legend cell align=left,align=left, draw=UCLblack},
]

\addplot+[color=UCLorange,solid,thick, mark=*]
  table[row sep=crcr]{
0000000010 0.0053648068669527905 \\ 
0000000020 0.01640656262505002 \\ 
0000000050 0.05965317919075145 \\ 
0000000100 0.08790874524714828 \\ 
0000000200 0.21694251646200394 \\ 
0000000500 0.5720194209674518 \\ 
0000001000 0.9253666954270923 \\ 
0000002000 1.0182898595462262 \\ 
0000005000 1.6840501701112072 \\ 
0000010000 2.0825910931174088 \\ 
0000020000 2.3703723385411135 \\ 
0000050000 4.088638911909521 \\ 
0000100000 5.057445336701856 \\ 
0000200000 5.362341910682023 \\ 
0000500000 7.748502595042424 \\ 
0001000000 7.945548348813635 \\ 
0002000000 8.811358829186311 \\ 
0005000000 9.182582433724049 \\ 
0010000000 9.130368687630796 \\ 
0015000000 8.059987730857165 \\ 
};
\addlegendentry{Intel i7-6700HQ};

\addplot+[color=UCLdarkBlue,solid, thick, mark=*]
  table[row sep=crcr]{
0000000010 0.000928993899606726 \\ 
0000000020 0.00264960579035802 \\ 
0000000050 0.008327147145208663 \\ 
0000000100 0.018478851625691357 \\ 
0000000200 0.03777151179004121 \\ 
0000000500 0.08907868944273313 \\ 
0000001000 0.15342615993514855 \\ 
0000002000 0.23258888438419914 \\ 
0000005000 0.383931465617041 \\ 
0000010000 0.5682818634532693 \\ 
0000020000 0.7157290630381994 \\ 
0000050000 1.1570145609303546 \\ 
0000100000 1.5513885044745825 \\ 
0000200000 2.045753360600733 \\ 
0000500000 4.116677225194478 \\ 
0001000000 5.743631141088108 \\ 
0002000000 8.75259877280893 \\ 
0005000000 12.103237479264495 \\ 
0010000000 15.084595208304782 \\ 
0015000000 16.686884818589057 \\ 
0020000000 18.41930051945805 \\ 
0050000000 21.140874951107055 \\ 
0100000000 23.363673366947687 \\ 
0150000000 24.351533928031678 \\ 
};
\addlegendentry{Intel Xeon Phi 7210};

\addplot+[color=UCLdarkRed, solid, thick, mark=*]
  table[row sep=crcr]{
10000 1.1602090592334493 \\ 
20000 1.0471596244131456 \\ 
50000 1.7542677824267783 \\ 
100000 3.006501340482574 \\ 
200000 3.9959703264094952 \\ 
500000 7.368420017482517 \\ 
1000000 10.134115615615615 \\ 
2000000 15.637247568318665 \\ 
5000000 24.246586276198677 \\ 
10000000 31.31806227411732 \\ 
15000000 37.64892395663152 \\ 
20000000 40.27012896116273 \\ 
50000000 41.23614804878049 \\ 
100000000 64.39806845663144 \\ 
200000000 65.16187417518375 \\ 
500000000 57.3363514076987 \\ 
1000000000 36.498524009326026 \\ 
};
\addlegendentry{AMD EPYC 7551};

\end{axis}
\end{tikzpicture}
	\caption{Number of tetrahedra created per second by our parallel implementation for different number of points. Tetrahedra are created more quickly when there is a lot of points because the proportion of conflicts is lower. An average rate of $65$ million tetrahedra created per second is obtained on the EPYC.}
	\label{Mtet_per_sec}
\end{minipage}
\end{figure}

\subsection{Ensuring termination}\label{subsec:reduction}

When the number of vertices of the current Delaunay triangulation is small, typically at the first steps 
of the algorithm, the probability that the mesh vertices 
belong to different partitions is very high and none of the point insertions may succeed. 
To avoid wasting precious milliseconds, the first BRIO round is always inserted sequentially.

Moreover, there is no guarantee that re-partitioning will be sufficient to insert all the points.
And even when a large part of the triangulation has already been constructed, parallel insertion may enter an infinite failure loop.
After a few rounds, the remaining points to insert are typically located in restricted volumes (intersection of previous buffer zones).
Because re-partitioning is done on the not-yet-inserted points, the resulting partitions are also small and thin.
This leads to inevitable conflicts as partitions get smaller than cavities.
In practice, we observed that the ratio $\rho$ of successful insertions decreases for a constant number of threads. 
If $80$ out of $100$ vertices are successfully inserted in a mesh ($\rho_k=0.8$), less that $16$ of the $20$ 
remaining vertices will be inserted at the following attempt ($\rho_{k+1}<0.8$).
Note that this effect is more important for small meshes, because the bigger the mesh, the relatively smaller the buffer zone and the higher the insertion success rate. This difference explains the growth of the number of tetrahedra created per second 
with the number of points in the triangulation (Figure~\ref{Mtet_per_sec}).

To avoid losing a significant amount of time in re-partitioning and trying to insert the same vertex multiple times
we gradually decrease the number of threads.
Choosing the adequate number of threads is a question of balance between the potential parallelization gain and
the partitioning cost that comes with each insertion attempt.
When decreasing the number of threads, we decrease the number of attempts needed.
When the ratio of successful insertions is too low, $\rho<1/5$, or when the number of points to insert per thread 
is under $3000$, we divide the number of threads by two.
Furthermore, if $\rho_k \le 1/n_{threads}$, the next insertion attempt will not benefit from multi-threading and we 
insert the points sequentially.

In Table~\ref{table:insertion} are given the number of threads used at each step of the 
Delaunay triangulation of a million vertices depending on the number of points to insert, the size of the 
current mesh, and the success insertion ratio $\rho$. Note that the computation of Moore indices and the 
sorting of vertices to insert are always computed on the maximal number of threads available ($8$ threads in 
this case) even when the insertion runs sequentially.

\begin{table}[htb]
    \centering
    \begin{tabular}{l r r r  c r}
        \toprule
                    & \multicolumn{3}{c}{\#points}  &  \#threads &\#mesh \\
					\cline{2-4}
                    & to insert & inserted & $\rho$  &   & vertices \\

        \midrule
        Initial mesh & & & & & 4 \\
        \midrule
        BRIO Round 1 & $2044$ & $2044$ & $100\%$ & $1$ & $2048$ \\
        \hline
        BRIO Round 2 & $12\,288$ & $6988$ & $57\%$ & $4$ & $9036$ \\
                     & $5300$ & $3544$ & $67\%$ & $2$ & $12\,580$ \\
                     & $1756$ & $1756$ & $100\%$ & $1$ & $14\,336$ \\
        \hline
        BRIO Round 3 & $86\,016$ & $59\,907$ & $70\%$ & $8$ & $74\,243$\\
                     & $26\,109$ & $11\,738$ & $45\%$ & $8$ & $85\,981$\\
                     & $14\,371$ & $7092$ & $49\%$ & $4$ & $93\,073$ \\
                     & $7279$  & $5332$ & $73\%$ & $2$ & $98\,405$ \\
                     & $1947$  & $1947$ & $100\%$ & $1$ & $100\,352$ \\
        \hline
        BRIO Round 4 & $602\,112$ & $503\,730$ & $84\%$ & $8$ & $604\,082$ \\
                     & $98\,382$  & $44\,959 $ & $46\%$ & $8$ & $649\,041$ \\
                     & $53\,423$  & $31\,702 $ & $59\%$ & $8$ & $680\,743$ \\
                     & $21\,721$  & $7903  $ & $36\%$ & $8$ & $688\,646$ \\
                     & $13\,818$  & $9400  $ & $68\%$ & $4$ & $698\,046$ \\
                     & $4418$   & $3641  $ & $82\%$ & $2$ & $701\,687$ \\
                     & $777$    & $777   $ & $100\%$ & $1$ & $702\,464$ \\
        \hline
        BRIO Round 5 & $297\,536$ & $271\,511$ & $91\%$ & $8$ & $973\,975$ \\
                     & $26\,025$  & $ 16\,426$ & $63\%$ & $8$ & $990\,401$ \\
                     & $9599$   & $  8092$ & $84\%$ & $4$ & $998\,493$ \\
                     & $1507$   & $  1507$ & $100\%$ & $1$ & $1\,000\,000$ \\
        \bottomrule \\
    \end{tabular}
    \caption{Numbers of threads used to insert points in our parallel Delaunay triangulation implementation 
	according to the number of points to insert, the mesh size and the insertion success at the previous step. $94.5\%$ of points are inserted using $8$ threads and $5\%$ using $4$ threads}
    \label{table:insertion}
\end{table}

\subsection{Data structures}

The data structure for our parallel Delaunay triangulation algorithm is similar to the one used by our sequential implementation (Section~\ref{sec:datastructure}).
There are two small differences. First, the parallel implementation does not compute sub-determinants for each tetrahedron. 
Actually, the bandwidth usage with the parallel insertions is already near its maximum. 
Loading and storing sub-determinant becomes detrimental to the overall performance. 
Instead we store a 16-bit\footnote{An 8-bit char would also work (not less or it would create data races) but the color flag is also used to distinguish volumes in our mesh generator described in \S\ref{sec:mesh_generation}} color flag to mark deleted tetrahedra. 
Second, each thread has its own \texttt{Cavity\_t} structure (Listing \ref{cavity_struct})
to which are added two integers identifying the starting and ending Moore indices of the thread's partition.

\subparagraph{Memory footprint}
Because we do not store four sub-determinants per tetrahedra anymore but only a 2-bytes color, our mesh data structure is lighter.
Still assuming that there is approximately $6n$ tetrahedra for $n$ vertices, it requires a little more than $6 \times 50 + 32 = 332$ bytes per vertex.
Thanks to this low memory footprint, we were able to compute the tetrahedralization of $N=10^9$ vertices (about 6 billion of
tetrahedra) on an AMD EPYC machine that has $512$ GB of RAM.
The experimental memory usage shown in Table \ref{fig:hxt_parallel_memory} differ slightly from the theoretical formula because (i) more memory is allocated than what is used (ii) measurements represent the maximum memory used by the whole program, including the stack, the text and data segment etc.

\begin{table}[htb]
\centering
    \begin{tabular}{L{1.4cm} r r r r}
    \toprule
    \# vertices & $10^4$ & $10^5$ & $10^6$ & $10^7$ \\
    \midrule
    \textcolor{UCLdarkBlue!80!black}{Ours} & $6.9$ \textsc{mb} & $43.8$ \textsc{mb} & $404.8$ \textsc{mb} & $3.8$ \textsc{gb} \\
    \textcolor{UCLorange!80!black}{Geogram} & $6.7$ \textsc{mb} & $30.5$ \textsc{mb} & $268.6$ \textsc{mb} & $2.7$ \textsc{gb}\\
    \textcolor{UCLgreen!80!black}{CGAL} & $14.1$ \textsc{mb} & $66.7$ \textsc{mb} & $578.8$ \textsc{mb} & $5.7$ \textsc{gb}\\
    \bottomrule \\
    \end{tabular}
    \caption{Comparison of the maximum memory usage of our parallel implementation, CGAL\cite{cgal:pt-t3-18a} and Geogram\cite{levy2015geogram} when using 8 threads.}
    \label{fig:hxt_parallel_memory}
\end{table}


\subsection{Critical operations}\label{subsec:critical}

When creating new tetrahedra in the Delaunay ball of a point, a thread first recycles unused memory space by replacing deleted tetrahedra. The indices of deleted tetrahedra are stored in the \texttt{cavity->deleted.tetrahedra\_ID}\ref{mesh_struct} array of each thread.
When the \texttt{cavity->deleted.tetrahedra\_ID} array of a thread is empty, 
additional memory should be reserved by this thread to create new tetrahedra.

This operation is a critical part of the program requiring synchronization between threads 
to respect the rule \ref{enum:rule2}. 
We need to capture the current number of tetrahedra and increment it
by the requested number of new tetrahedra in one single atomic operation. 
\emph{OpenMP} provides the adequate mechanism, see Listing \ref{asking_tetrahedra_code}.

To reduce the number of time this operation is done, the number of tetrahedra is incremented atomically by at least $8192$, and the \texttt{cavity->deleted.tetrahedra\_ID} is filled with the new indices of tetrahedra. Those tetrahedra are conceptually deleted although they have never been in the mesh.
The number $8192$ was found experimentally among multiples of 
$512$\footnote{It is common to choose a multiple of the page size (usually $4096$ bytes) to minimize TLB misses.}.
Increasing it would reduce the number of time reservation of new tetrahedra is performed but it is not necessary.
Indeed, since the critical operation occurs then in average every $1000^+$ insertions, 
the time wasted is very low.
Therefore, increasing the default number of deleted tetrahedra would only wastes memory space for little to no gain.

When the space used by tetrahedra exceeds the initially allocated capacity, 
reallocation is implemented that doubles the capacity 
of the arrays of mesh tetrahedra.
During that operation, memory is potentially moved at another location. 
No other operation can be performed at that time on the mesh. 
Therefore, the reallocation code is placed in between two OpenMP barriers (Listing \ref{reallocTetrahedraIfNeeded}). 
This synchronization event is very rare and does not impact performances.
In general, a good estimation of the needed memory needed to reserve is possible and  
that critical section is never reached.

\begin{lstlisting}[caption={When there are less deleted tetrahedra than there are tetrahedra in the Delaunay ball,
 $8192$ new "deleted tetrahedra" indices are reserved by the thread. As the \texttt{mesh} data structure is shared by all threads,
 \texttt{mesh->tetrahedra.num} must be increased in one single atomic operation.
 %This is the only portion of the Delaunay insertion that include synchronization mechanisms.
 }, label=asking_tetrahedra_code, captionpos=b, float]
if(cavity->to_create.num > cavity->deleted.num)
{
    uint64_t nTetNeeded = MAX(8192, cavity->to_create.num) - cavity->deleted.num;

    uint64_t nTet;
    #pragma omp atomic capture
    {
        nTet = mesh->tetrahedra.num;
        mesh->tetrahedra.num+=nTetNeeded;
    }

    reallocTetrahedraIfNeeded(mesh);
    reallocDeletedIfNeeded(state, cavity->deleted.num + nTetNeeded);

    for (uint64_t i=0; i<nTetNeeded; i++){
        cavity->deleted.tetrahedra_ID[cavity->deleted.num+i] = 4*(nTet+i);
        mesh->tetrahedra.color[nTet+i] = DELETED_COLOR;
    }

    cavity->deleted.num += nTetNeeded;
}
\end{lstlisting}

\begin{lstlisting}[caption={Memory allocation for new tetrahedra is synchronized with OpenMP barriers.}, label=reallocTetrahedraIfNeeded, captionpos=b, float]
void reallocTetrahedraIfNeeded(mesh_t* mesh)
{
    if(mesh->tetrahedra.num > mesh->tetrahedra.allocated_num)
    {
        #pragma omp barrier

        // all threads are blocked except the one doing the reallocation
        #pragma omp single
        {
            uint64_t nTet = mesh->tetrahedra.num;
            alignedRealloc(&mesh->tetrahedra.neighbor_ID, nTet*8*sizeof(uint64_t));
            alignedRealloc(&mesh->tetrahedra.vertex_ID, nTet*8*sizeof(uint32_t));
            alignedRealloc(&mesh->tetrahedra.color, nTet*2*sizeof(uint16_t));
            mesh->tetrahedra.allocated_num = 2*nTet;

        } // implicit OpenMP barrier here
    }
}
\end{lstlisting}


\subsection{Parallel implementation performances}\label{sec:parrallel_performances}

We are able to compute the Delaunay triangulation of over one billion tetrahedra in record-breaking time:
41.6 seconds on the Intel Xeon Phi and 17.8 seconds on the AMD EPYC.
These timings do not include I/Os.
As for the title of this article, we are able to generate three billion tetrahedra in 53 seconds on the EPYC. 
The scaling of our implementation regarding the number of threads is detailed in Figure \ref{fig:hxt_parallel_scaling}.
We obtain a good scaling until the number of threads reach the number of cores, 
 i.e. 4 cores for the Intel i7-6700HQ, 64 cores for the Intel Xeon Phi 7210 and the AMD EPYC 7551.
To our best knowledge, CGAL\cite{cgal:pt-t3-18a} and Geogram\cite{levy2015geogram} are the two fastest open-source CPU implementations available for 3D Delaunay triangulation.
We compare their performances to ours on a laptop
(Figure~\ref{fig:hxt_parallel_i7}) and on the Intel Xeon Phi (Figure~\ref{fig:hxt_parallel_KNL}). 


\begin{figure}[!htb]
\begin{subfigure}[b]{\textwidth}
\begin{subfigure}[b]{0.58\textwidth}
    \setlength\figureheight{0.6\textwidth}
    \setlength\figurewidth{\textwidth}
    \begin{tikzpicture}

\begin{axis}[
width=\figurewidth,
height=\figureheight,
xmode=log,
xmin=10000,
xmax=10000000,
xlabel={Number of points (random uniform distribution)},
ymode=log,
ymin=0.03,
ymax=25,
ylabel={Time [s]},
log y ticks with fixed point,
]

\addplot
  table[row sep=crcr]{
0000000010 0.005592 \\
0000000020 0.004998 \\
0000000050 0.004325 \\
0000000100 0.006575 \\
0000000200 0.005619 \\
0000000500 0.005561 \\
0000001000 0.006954 \\
0000002000 0.012958 \\
0000005000 0.019693 \\
0000010000 0.032110 \\
0000020000 0.056454 \\
0000050000 0.082052 \\
0000100000 0.132996 \\
0000200000 0.251282 \\
0000500000 0.435253 \\
0001000000 0.849359 \\
0002000000 1.532746 \\
0005000000 3.679653 \\
0010000000 7.402391 \\
};
\addlegendentry{Ours};

\addplot
  table[row sep=crcr]{
0000000010 0.020000 \\
0000000020 0.020000 \\
0000000050 0.020000 \\
0000000100 0.020000 \\
0000000200 0.020000 \\
0000000500 0.020000 \\
0000001000 0.020000 \\
0000002000 0.021500 \\
0000005000 0.040500 \\
0000010000 0.040500 \\
0000020000 0.070500 \\
0000050000 0.143500 \\
0000100000 0.190000 \\
0000200000 0.363333 \\
0000500000 0.913333 \\
0001000000 1.730000 \\
0002000000 3.370000 \\
0005000000 8.350000 \\
0010000000 17.106667 \\
};
\addlegendentry{Geogram 1.5.4};

\addplot
  table[row sep=crcr]{
0000000010 0.000639 \\
0000000020 0.000668 \\
0000000050 0.000819 \\
0000000100 0.001166 \\
0000000200 0.005489 \\
0000000500 0.005936 \\
0000001000 0.008117 \\
0000002000 0.012515 \\
0000005000 0.021290 \\
0000010000 0.036762 \\
0000020000 0.058673 \\
0000050000 0.126478 \\
0000100000 0.237791 \\
0000200000 0.460530 \\
0000500000 1.136561 \\
0001000000 2.198305 \\
0002000000 4.651433 \\
0005000000 11.967902 \\
0010000000 23.374591 \\
};
\addlegendentry{CGAL 4.12};


\end{axis}
\end{tikzpicture}
\end{subfigure}
~
\begin{subfigure}[b]{0.38\textwidth}
    \begin{tabular}{L{1.4cm} r r r r}
    \toprule
    \# vertices & $10^4$ & $10^5$ & $10^6$ & $10^7$ \\
    \midrule
    \textcolor{UCLdarkBlue!80!black}{Ours} & 0.032 & 0.13 & 0.85 & 7.40 \\
    \textcolor{UCLorange!80!black}{Geogram} & 0.041 & 0.19 & 1.73 & 17.11\\
    \textcolor{UCLgreen!80!black}{CGAL} & 0.037 & 0.24 & 2.20 & 23.37\\
    \bottomrule
    \end{tabular}
    \vspace{1.5cm}
\end{subfigure}
    \caption{4-core Intel$^\circledR$ Core$^\text{TM}$ i7-6700HQ CPU.}
    \label{fig:hxt_parallel_i7}
\end{subfigure}

\vspace{0.5cm}
\begin{subfigure}[b]{\textwidth}
\begin{subfigure}[b]{0.58\textwidth}
    \setlength\figureheight{0.6\textwidth}
    \setlength\figurewidth{\textwidth}
    \begin{tikzpicture}

\begin{axis}[
width=\figurewidth,
height=\figureheight,
xmode=log,
xmin=10000,
xmax=150000000,
xlabel={Number of points (random uniform distribution)},
ymode=log,
ymin=0.1,
ymax=230,
ylabel={Time [s]},
log y ticks with fixed point,
]

\addplot
  table[row sep=crcr]{
0000000010 0.032293 \\
0000000020 0.030948 \\
0000000050 0.030983 \\
0000000100 0.031279 \\
0000000200 0.032273 \\
0000000500 0.035710 \\
0000001000 0.041942 \\
0000002000 0.056731 \\
0000005000 0.086380 \\
0000010000 0.117674 \\
0000020000 0.186966 \\
0000050000 0.289954 \\
0000100000 0.433560 \\
0000200000 0.658662 \\
0000500000 0.819243 \\
0001000000 1.174975 \\
0002000000 1.543036 \\
0005000000 2.791709 \\
0010000000 4.480502 \\
0015000000 6.076866 \\
0020000000 7.341305 \\
0050000000 15.994128 \\
0100000000 28.950726 \\
0150000000 41.666166 \\
};
\addlegendentry{Ours};

\addplot
  table[row sep=crcr]{
0000000010 0.020000 \\
0000000020 0.020000 \\
0000000050 0.020000 \\
0000000100 0.020000 \\
0000000200 0.020000 \\
0000000500 0.030000 \\
0000001000 0.040000 \\
0000002000 0.070000 \\
0000005000 0.150000 \\
0000010000 0.103333 \\
0000020000 0.176667 \\
0000050000 0.400000 \\
0000100000 0.536667 \\
0000200000 1.046667 \\
0000500000 2.596667 \\
0001000000 4.580000 \\
0002000000 9.306667 \\
0005000000 22.056667 \\
0010000000 43.703333 \\
0020000000 89.320000 \\
0050000000 229.426667 \\
};
\addlegendentry{Geogram 1.5.4};

\addplot
  table[row sep=crcr]{
0000000010 0.000405 \\
0000000020 0.000615 \\
0000000050 0.001238 \\
0000000100 0.002506 \\
0000000200 0.038967 \\
0000000500 0.115835 \\
0000001000 0.164683 \\
0000002000 0.198776 \\
0000005000 0.223968 \\
0000010000 0.272806 \\
0000020000 0.292985 \\
0000050000 0.406505 \\
0000100000 0.481568 \\
0000200000 0.692698 \\
0000500000 1.382804 \\
0001000000 2.439070 \\
0002000000 4.275555 \\
0005000000 10.779064 \\
0010000000 20.153834 \\
0020000000 40.428774 \\
0050000000 102.462438 \\
};
\addlegendentry{CGAL 4.12};


\end{axis}
\end{tikzpicture}
\end{subfigure}
\begin{subfigure}[b]{0.41\textwidth}
    \begin{tabular}{L{1.4cm} r r r r r}
    \toprule
    \# vertices & $10^4$ & $10^5$ & $10^6$ & $10^7$ & $10^8$\\
    \midrule
    \textcolor{UCLdarkBlue!80!black}{Ours} & 0.11 & 0.43 & 1.17 & 4.48 & 28.95 \\
    \textcolor{UCLorange!80!black}{Geogram} & 0.10 & 0.54 & 4.58 & 43.70 & /~~~\\
    \textcolor{UCLgreen!80!black}{CGAL} & 0.27 & 0.48 & 2.44 & 20.15 & /~~~\\
    \bottomrule
    \end{tabular}
    \vspace{1.5cm}
\end{subfigure}
    \caption{64-core Intel$^\circledR$ Xeon Phi$^\text{TM}$ 7210 CPU.}
    \label{fig:hxt_parallel_KNL}
\end{subfigure}
\caption[]{Comparison of our parallel implementation with the parallel implementation in CGAL\cite{cgal:pt-t3-18a} and Geogram\cite{levy2015geogram} with on a high-end laptop \textbf{(a)} and a many-core computer \textbf{(b)}.
All timings are in seconds.
}
\end{figure}

\section{Tetrahedral mesh generation} \label{sec:mesh_generation}

The parallel Delaunay triangulation algorithm that we presented 
in the previous section is integrated in a Delaunay refinement mesh generator.
A tetrahedral mesh generator is a procedure that takes as 
input the boundary of a domain to mesh, defined 
by set of triangles $t$ that defines the boundary of a closed volume, 
and that returns a finite element mesh, i.e. a set of tetrahedra $\mathcal T$ of controlled
sizes and shapes which boundary is equal to the input triangulation: $\partial \mathcal T = t$.
From a coarse mesh that is conformal to $t$, a Delaunay-refinement based mesher inserts progressively vertices 
until element sizes and shapes follow the prescribed ranges.
Generating a mesh is an intrinsically more difficult problem than constructing 
the Delaunay triangulation of given points because 
(1) the points are not given, 
(2) the boundary triangles must be facets of the generated tetrahedra, and 
(3)  the tetrahedra shape and size should be optimized to maximize the mesh quality.
Our objective in this section is to demonstrate that our parallel Delaunay point insertion 
may be integrated in a mesh generator.
The interested reader is referred to the book by Frey and George \cite{Frey:2007:MGA:1205626} for a complete review of finite-element mesh generation.


The mesh generation algorithm \ref{algo:mesher} proposed in this section 
follows the approach implemented for example in Tetgen\cite{DBLP:journals/toms/Si15}. 
First, all the vertices of the boundary triangulation $t$ are
tetrahedralized to form the initial ``empty mesh'' ${\mathcal  T}_0$. 
Then, ${\mathcal  T}_0$ is transformed into a conformal mesh
$\mathcal T$  which boundary $\partial {\mathcal T}$ 
is equal to $t$ : $\partial \mathcal T = t$.
The triangulation $\mathcal T$ is then progressively refined by (i) creating 
vertices $\pointSet$ at the circumcenters of tetrahedra for which the circumsphere radius, $r_{\tau}$, 
is significantly larger than the desired mesh size, $h$, i.e. $r_{\tau} / h > 1.4$ \cite{shewchuk1997delaunay}. 
(ii) inserting the vertices in the mesh using our parallel Delaunay algorithm.
The sets of points $\pointSet$ to insert are filtered {\it a priori} in order not to generate short edges,
i.e. edges of size smaller than $0.7 h$.
We first use Hilbert coordinates to discard very close points on
the curve, and implemented a slightly modified cavity algorithm that aborts if a point of the cavity is too close from the one to insert (Section~\ref{sec:cavity}).
Points are thus discarded both in the filtering process and in the insertion process. 
Note that contrary to existing implementations, we insert as large as possible point sets $\pointSet$ 
during the refinement step to take advantage of the efficiency of our parallel point insertion algorithm (Section~\ref{sec:parallel}).

We parallelized all steps of the mesh generator (Algorithm \ref{algo:mesher}) 
except the boundary recovery procedure that is the one of Gmsh \cite{geuzaine_gmsh:_2009} which is
essentially based on Tetgen \cite{DBLP:journals/toms/Si15}.

The cavity algorithm has also been modified to accommodate possible constraints on the faces and edges of tetrahedra. 
With this modification, mesh refinement never breaks the surface mesh and constrained edges and faces can be included in the mesh interior.
As a result, the mesh may not satisfy the Delaunay property anymore.
Therefore, we must always ensure that cavities are star-shaped. This is achieved by a simple procedure that checks if boundary facets of the cavity are oriented such that they form a positive volume with the new point. If a boundary facet is not oriented correctly, indicating that the cavity is not star-shaped, the tetrahedron containing it is removed from the cavity and the procedure is repeated.
In practice, these modifications do not affect the speed of point insertions significantly.

To generate high-quality meshes, an optimization step should be added to the algorithm
to improve the general quality of the mesh elements and remove sliver tetrahedra.
Mesh improvements are out of the scope of this paper. However, 
first experiments have shown that optimization step should slow down the mesh generation by a factor two. 
For example, mesh refinement and improvement take approximately the same time in Gmsh\cite{geuzaine_gmsh:_2009}.

\begin{algorithm}
\caption{Mesh generation algorithm} 
\label{algo:mesher}
\textbf{Input:} A set of triangles $t$\\
\textbf{Output:} A tetrahedral mesh $\mathcal T$
\begin{algorithmic}[1]
    \Function{Parallel Mesher}{$t$}
		\State ${\mathcal T}_0 \gets$ {\sc{EmptyMesh}}($t$)
		\State ${\mathcal T}    \gets$ {\sc{RecoverBoundary}}(${\mathcal T}_0$)

        \While {${\mathcal T}$ contains large tetrahedra}
            \State $\pointSet \gets$     \Call{SamplePoints}{${\mathcal T}$}
            \State $\pointSet \gets$     \Call{FilterPoints}{${\pointSet}$}
            \State ${\mathcal T} \gets$ \Call{InsertPoints}{${\mathcal T}$, $\pointSet$}
        \EndWhile

        \State \Return ${\mathcal T}$
    \EndFunction
\end{algorithmic}
\end{algorithm}

\subsection {Small and medium size test cases on standard laptops}
\begin{figure}[h!]
\begin{subfigure}[l]{0.6\textwidth}
\begin{center}
\centering
\includegraphics[width=0.8\textwidth]{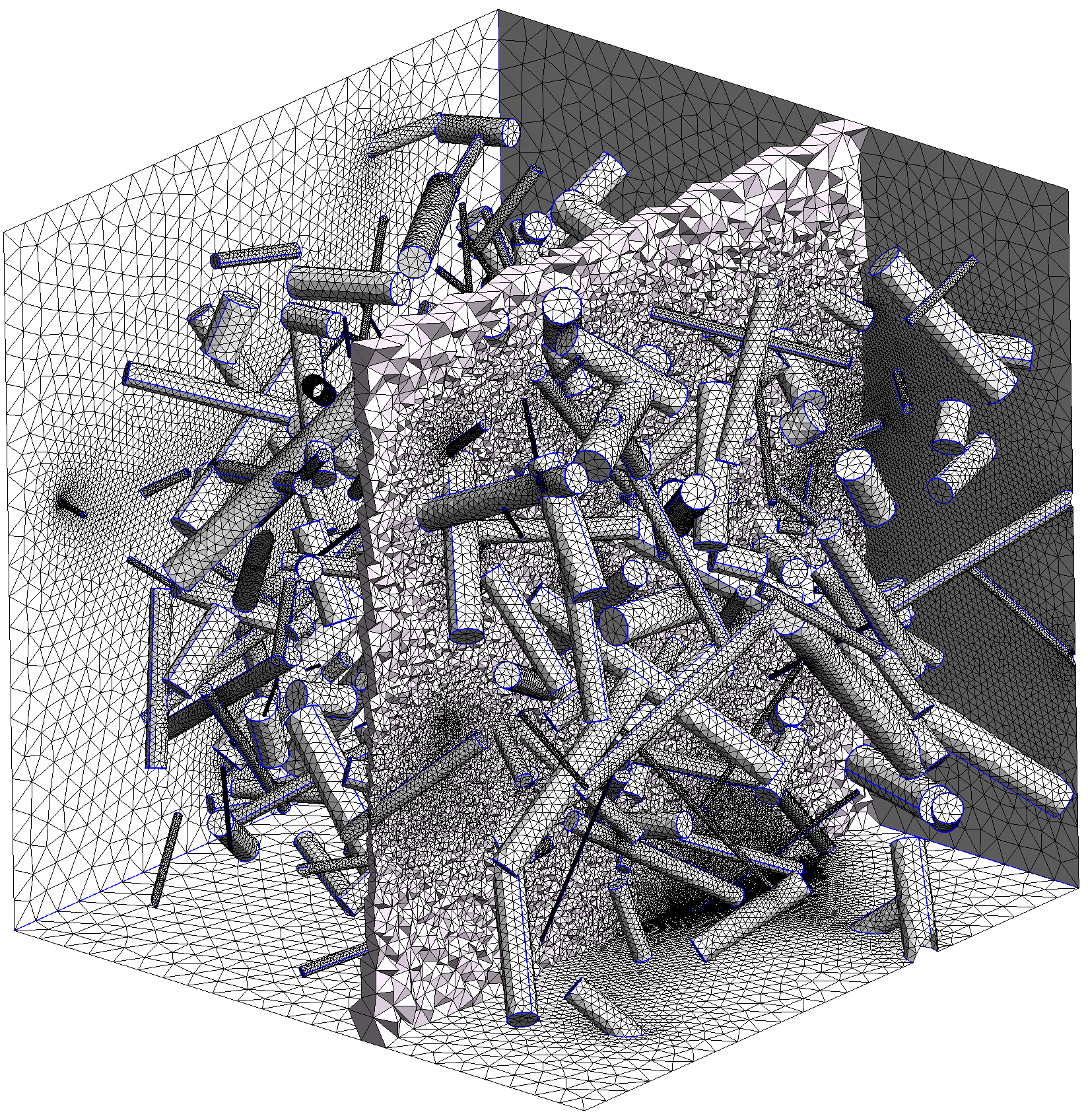}
\end{center}
\end{subfigure}
\begin{subfigure}[c]{0.4\textwidth}
\centering

\begin{tabular}{c}
 {\bf 100 fibers}
\end{tabular}

\begin{tabular}{cr rrr}
 \toprule
\multirow{2}{*}{\# threads} & \multirow{2}{*}{\# tetrahedra} & \multicolumn{3}{c}{Timings (s)} \\  
& & BR & Refine & Total  \\
\midrule
  1 &  12\,608\,242 & 0.74 & 19.6 & 20.8\\
  2 &  12\,600\,859 & 0.72 & 13.6 & 14.6\\
  4 &  12\,567\,576 & 0.72 & 8.7 & 9.8\\
  8 &  12\,586\,972 & 0.71 & 7.6 & 8.7\\
\bottomrule
\end{tabular}

\begin{tabular}{c}
\\
 {\bf 300 fibers}
\end{tabular}

\begin{tabular}{cr rrr}
\toprule
\multirow{2}{*}{\# threads} & \multirow{2}{*}{\# tetrahedra} & \multicolumn{3}{c}{Timings (s)} \\  
& & BR & Refine & Total  \\
\midrule
  1 &  52\,796\,891   & 6.03 & 92.4 & 101.3\\
  2 &  52\,635\,891   & 5.76 & 61.2 & 69.0 \\
  4 &  52\,768\,565   & 5.71 & 39.4 & 46.8 \\
  8 &  52\,672\,898   & 5.67 & 32.5 & 39.8 \\
\bottomrule
\end{tabular}
\end{subfigure}

\begin{subfigure}[l]{0.6\textwidth}
\begin{center}
\centering
\includegraphics[width=0.9\textwidth]{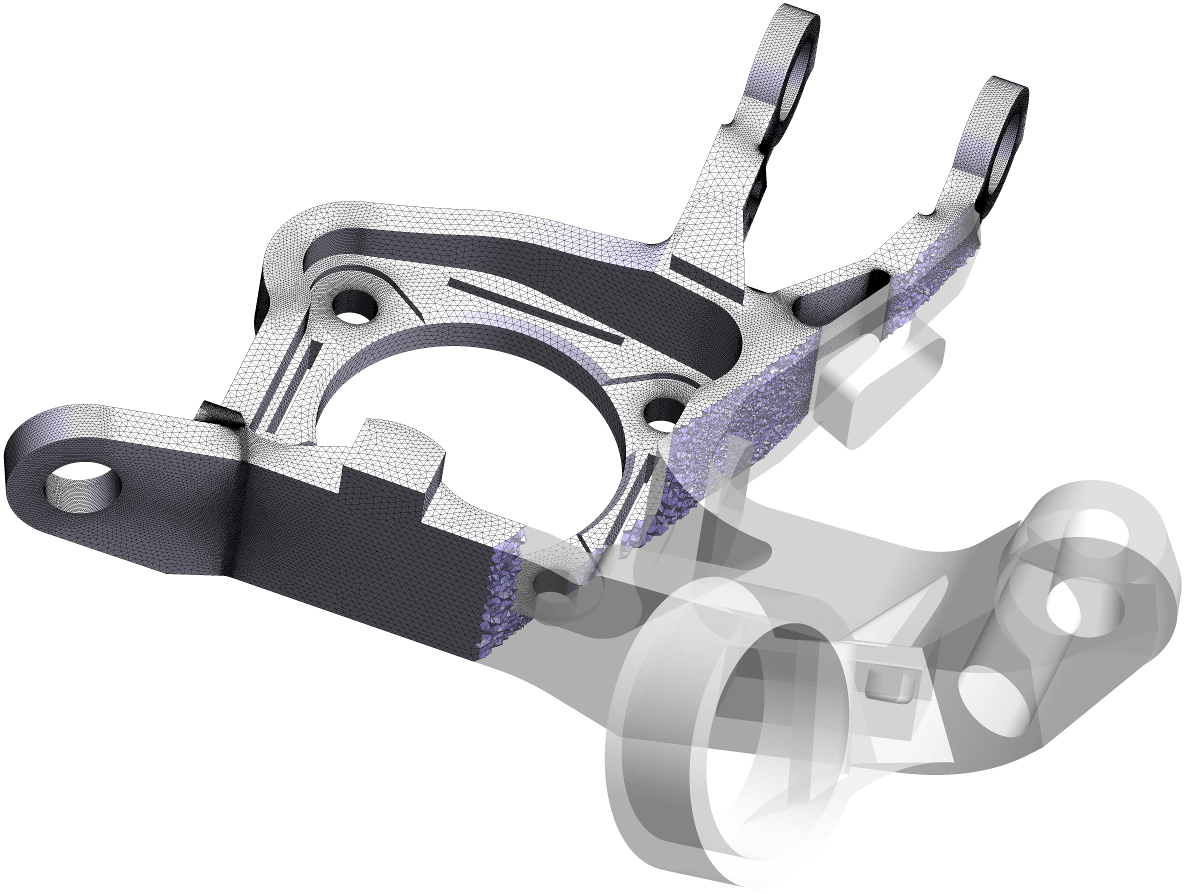}
\end{center}
\end{subfigure}
\begin{subfigure}[c]{0.4\textwidth}
\centering

\begin{tabular}{c}
 {\bf Mechanical part}
\end{tabular}

\begin{tabular}{cr rrr}
 \toprule
\multirow{2}{*}{\# threads} & \multirow{2}{*}{\# tetrahedra} & \multicolumn{3}{c}{Timings (s)} \\  
& & BR & Refine & Total  \\
\midrule
 1 & 24\,275\,207  & 8.6 & 43.6 & 56.3 \\
 2 & 24\,290\,299  & 8.4 & 30.4 & 41.8 \\
 4 & 24\,236\,112  & 8.1 & 24.6 & 35.3 \\
 8 & 24\,230\,468  & 8.1 & 21.8 & 32.6 \\
\bottomrule
\end{tabular}
\end{subfigure}

\begin{subfigure}[l]{0.6\textwidth}
\begin{center}
\centering
\includegraphics[width=0.99\textwidth]{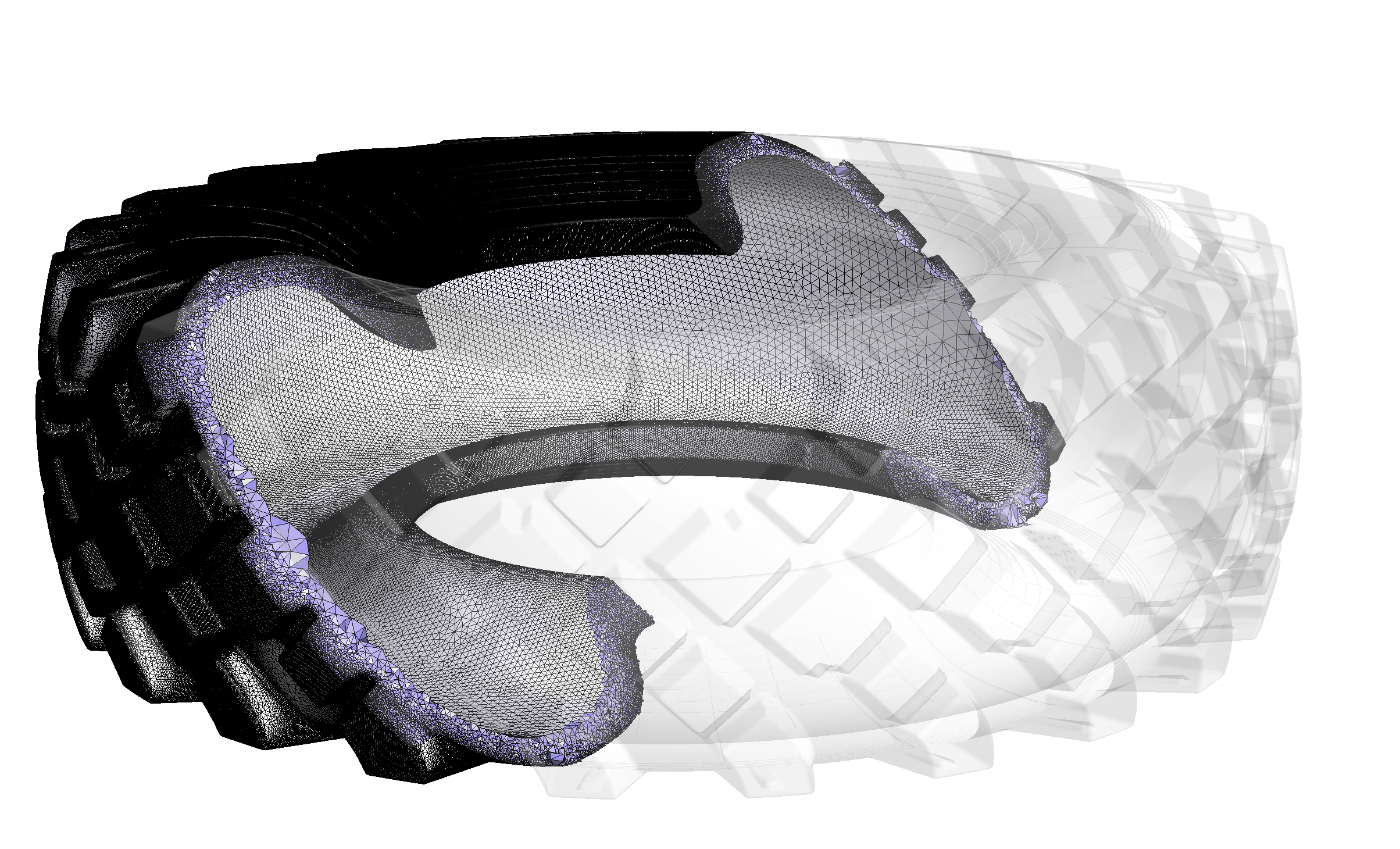}
\end{center}
\end{subfigure}
\begin{subfigure}[c]{0.4\textwidth}
\centering

\begin{tabular}{c}
 {\bf Truck tire}
\end{tabular}

\begin{tabular}{cr rrr}
 \toprule
\multirow{2}{*}{\# threads} & \multirow{2}{*}{\# tetrahedra} & \multicolumn{3}{c}{Timings (s)} \\  
& & BR & Refine & Total  \\
\midrule
 1 & 123\,640\,429  & 75.9 & 259.7 & 364.7 \\
 2 & 123\,593\,913   & 74.5 & 166.8 & 267.1 \\
 4 & 123\,625\,696  & 74.2 & 107.4 & 203.6 \\
 8 & 123\,452\,318  & 74.2 & 95.5 & 190.0 \\
\bottomrule
\end{tabular}
\end{subfigure}
\caption{Performances of our parallel mesh generator on a Intel$^\circledR$ Core$^\text{TM}$ i7-6700HQ 4-core 
CPU. Wall clock times are given for the whole meshing process for 1 to 8 threads. 
They include IOs (sequential), initial mesh generation (parallel), as well as sequential boundary recovery (BR),
and parallel Delaunay refinement for which detailed timings are given. \label{fig:small}}
\end{figure}

In order to verify the scalability of the whole meshing process, meshes of up to one hundred million tetrahedra were computed on a $4$ core 3.5 GHz Intel Core i7-6700HQ with
$1$, $2$, $4$ and $8$ threads. Those meshes easily fit within the 8Gb of RAM of this modern laptop.

Three benchmarks are considered in this section: (i)
a cube filled with cylindrical fibers of random radii and lengths
that are randomly oriented, (ii) a mechanical part and (iii) a truck tire.
Surface meshes are computed with Gmsh\cite{geuzaine_gmsh:_2009}. Mesh size on the
surfaces is controlled by surface curvatures and mesh size inside the domain 
is simply interpolated from the surface mesh. 

Illustrations of the meshes, as well as timings statistics are presented in Figure~\ref{fig:small}. 
Our mesher is able to generate between $40$ and $100$ 
million tetrahedra per minute.
Using multiple threads allows some speedup, the mesh refinement process is accelerated
by a factor ranging between $2$ and $3$ on this
$4$ core machine.

The last test case (truck tire) is defined by more than $7000$ CAD surfaces. 
Recovering the $27\,892$ triangular facets missing from $\mathcal T_0$ 
takes more than a third of the total meshing time with the maximal number of threads. 
Parallelizing the boundary recovery process is clearly a priority of our future developments.
On this same example, the surface mesh was done with Gmsh using four threads. 
The surface mesher of Gmsh is not very fast and it took about the same time to generate the surface
mesh of $6\,881\,921$ triangles as to generate the volume mesh that
contains over one hundred million tetrahedra using the same number of threads. 
The overall meshing time for the truck tire test case
is thus about $6$ minutes.




\subsection{Large mesh generation on many core machine}
We further generated meshes containing over 300,000,000 elements  on a AMD$^\circledR$ EPYC 64 core 
machine.  
Three benchmarks are considered: (i) two cubes filled with many  randomly oriented cylindrical fibers of random radii and lengths, 
and (ii) the exterior of an aircraft.
Surface meshes were also generated by Gmsh. 

Our strategy reaches its maximum efficiency for large meshes. In the
500 thin fibers test case, over 700,000,000 tetrahedra were generated
in 135 seconds. This represents a rate of 5.2 million tetrahedra per
second.  In the 500 thin fibers test case, boundary recovery cost was 
lower and a rate of 6.2 million tetrahedra per
second was reached.  

\begin{figure}[h!]
\begin{subfigure}[l]{0.6\textwidth}
\begin{center}
\centering
\includegraphics[width=0.8\textwidth]{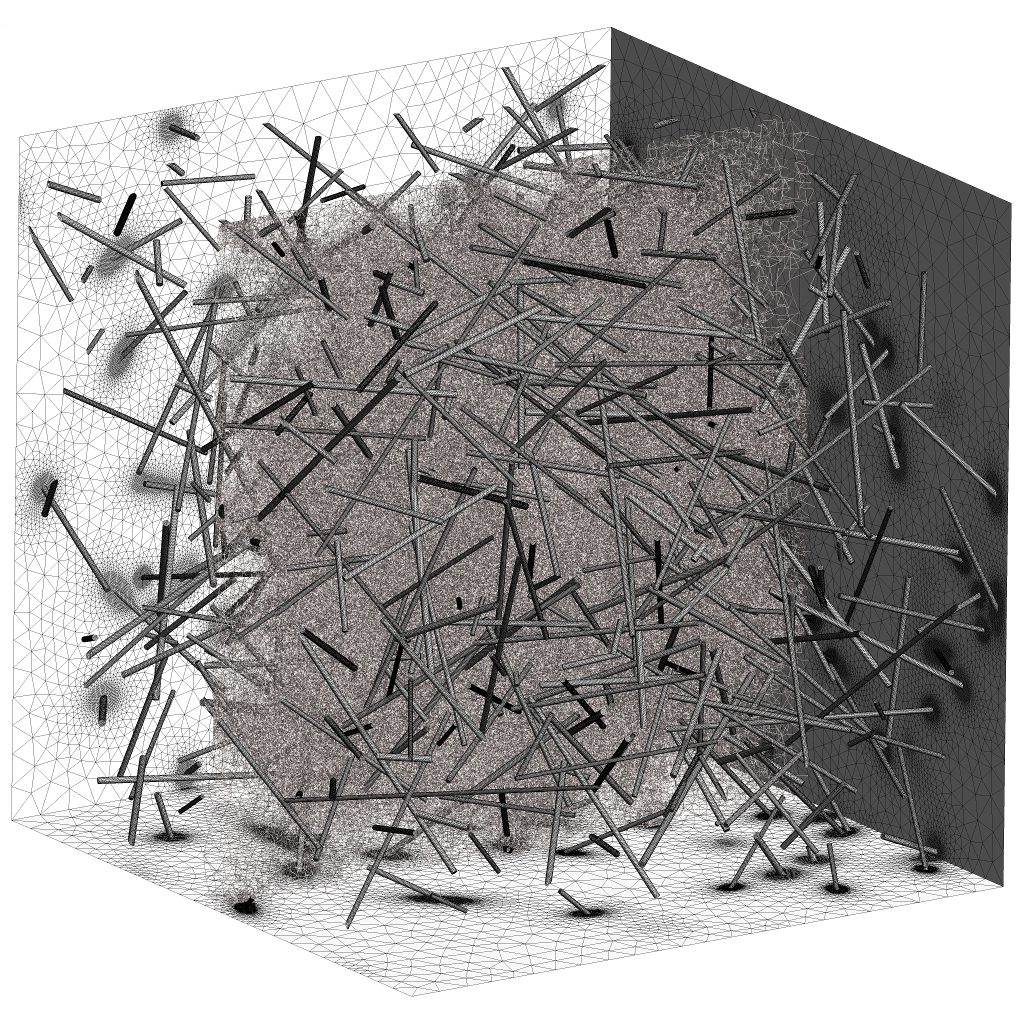}
\end{center}
\end{subfigure}
\begin{subfigure}[c]{0.4\textwidth}
\centering

\begin{tabular}{c}
 {\bf 100 thin fibers}
\end{tabular}

\begin{tabular}{cr rrr}
\toprule
\multirow{2}{*}{\# threads} & \multirow{2}{*}{\# tetrahedra} & \multicolumn{3}{c}{Timings (s)} \\ 
& & BR & Refine & Total  \\
\midrule
  1 &  325\,611\,841   & 3.1 & 492.1 & 497.2 \\
  2 &  325\,786\,170   & 2.9 &  329.7 & 334.3 \\
  4 &  325\,691\,796   & 2.8 & 229.5 & 233.9 \\
  8 &  325\,211\,989   & 2.7 & 154.6 & 158.7 \\
  16 &  324\,897\,471   & 2.8 & 96.8 & 100.9 \\
  32 &  325\,221\,244   & 2.7 & 71.7 & 75.8 \\
  64 &  324\,701\,883   & 2.8 & 55.8 & 60.1 \\
  127 &  324\,190\,447   & 2.9 & 47.6 & 52.0 \\
\bottomrule
\end{tabular}

\begin{tabular}{c}
\\
 {\bf 500 thin fibers}
\end{tabular}

\begin{tabular}{cr rrr}
\toprule
\multirow{2}{*}{\# threads} & \multirow{2}{*}{\# tetrahedra} & \multicolumn{3}{c}{Timings (s)} \\
& & BR & Refine & Total  \\
\midrule
  1 &  723\,208\,595   & 18.9 & 1205.8 & 1234.4\\
  2 &  723\,098\,577   & 16.0 & 780.3 & 804.8 \\
  4 &  722\,664\,991   & 86.6 & 567.1 & 659.8 \\
  8 &  722\,329\,174   & 15.8 & 349.1 & 370.1 \\
  16 &  723\,093\,143   & 15.6 & 216.2 & 236.5 \\
  32 &  722\,013\,476   & 15.6 & 149.7 & 169.8 \\
  64 &  721\,572\,235   & 15.9 & 119.7 & 140.4 \\
  127 &  721\,591\,846   & 15.9 & 114.2 & 135.2 \\
\bottomrule
\end{tabular}
\end{subfigure}

\begin{subfigure}[l]{0.6\textwidth}
\begin{center}
\centering
\includegraphics[width=0.9\textwidth]{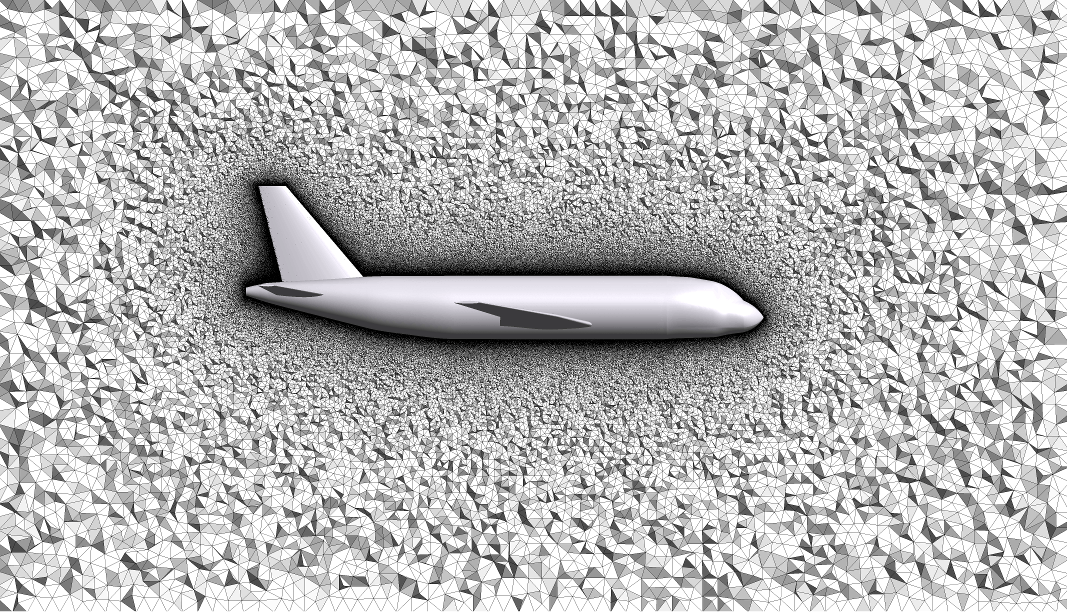}
\end{center}
\end{subfigure}
\begin{subfigure}[c]{0.4\textwidth}
\centering

\begin{tabular}{c}
 {\bf Aircraft}
\end{tabular}

\begin{tabular}{cr rrr}
 \toprule
\multirow{2}{*}{\# threads} & \multirow{2}{*}{\# tetrahedra} & \multicolumn{3}{c}{Timings (s)} \\  
 & & BR & Refine & Total  \\
\midrule
  1 &  672\,209\,630   & 45.2 & 1348.5 & 1418.3\\
  2 &  671\,432\,038   & 42.1 & 1148.9 & 1211.5 \\
  8 &  665\,826\,109   & 39.6 & 714.8 & 774.8 \\
  64 &  664\,587\,093   & 38.7 & 322.3 & 380.9  \\
  127 & 663\,921\,974   & 38.1 & 255.0 & 313.3 \\
\bottomrule
\end{tabular}
\end{subfigure}

\caption{Performances of our parallel mesh generator on a AMD$^\circledR$ EPYC 64-core 
machine. Wall clock times are given for the whole meshing process for 1 to 127 threads. 
They include IOs (sequential), initial mesh generation (parallel), as well as sequential boundary recovery (BR),
and parallel Delaunay refinement for which detailed timings are given. \label{fig:big}}

\end{figure}


\section{Conclusion}

This paper introduces a scalable Delaunay triangulation algorithm 
and demonstrates that inserting points concurrently 
can be performed effectively on shared memory architectures without the need of heavy synchronization mechanisms.
Our reference implementation is open-source and available in Gmsh 4 (www.gmsh.info).
We optimized both the sequential and the parallel algorithm by specifying them exclusively 
for Delaunay triangulation purposes.
Our parallel implementation of 3D Delaunay triangulation construction has one important drawback: 
partitioning is more costly for non-uniform point sets because Moore indices are not adaptive.
Adaptive Hilbert or Moore curve computation is relatively simple\cite{DBLP:journals/ipl/HamiltonR08, su_rapid_2016} 
but may not be faster than our refinement and sorting in one batch approach.
Another drawback of our implementation is that it might not be well suited for distributed memory architectures.
However, nothing prevents our strategy from being included in any larger divide-and-conquer approach.

We additionally presented a lightweight mesh generator based on Delaunay refinement.
This mesh generator is almost entirely parallel and is able to produce computational meshes with a rate above 5 million tetrahedra 
per second.
One issue when parallelizing mesh generation is to ensure reproducibility.
For mesh generation, it is usually admitted
that mesh generation should be reproducible for a given number of threads.
Our mesh generator has been made reproducible in that sense, 
with a loss of 20\% in overall performance. 
This feature has been implemented as an option.
On the other hand, Figures \ref{fig:small} and \ref{fig:big} show that, 
starting from the same input, the number of elements varies.
Making the code reproducible independently of the number of threads would dramatically harm its scalability.

Note that our mesh generator does not include the final optimization step of the mesh, \emph{mesh improvement},
to obtain a mesh adequate for finite-element simulations.
This crucial step of any industrial-grade meshing procedure is one major limitation of our 
mesh generator.
The basic operations to perform:  edge swap, edge collapse, edge split, vertex relocation and face swap operations will 
be done using the same parallel framework than the one we described to build the Delaunay triangulation.
The final challenge is the parallelization of the boundary recovery procedure.


\section*{Acknowledgments}
\vspace*{-5pt}
This project has received funding from the European Research Council (ERC) under the European Union's Horizon 2020 
research and innovation programme (grant agreement ERC-2015-AdG-694020).

\bibliography{biblio}
 
\end{document}